\documentclass[aps, prx, twocolumn,shortbibliography,natbib=true,url=false]{revtex4-2}
\usepackage{amsmath,amssymb}
\usepackage{graphicx}
\usepackage{color}
\usepackage[dvipsnames]{xcolor}
\usepackage{hyperref}
\usepackage{blindtext}
\usepackage[T1]{fontenc}
\usepackage[utf8]{inputenc}
\usepackage{multirow}
\usepackage{rotating}
\usepackage{lipsum}
\usepackage{mathtools}
\usepackage{tabularx}
\usepackage{graphicx}
\usepackage{dcolumn}
\usepackage{bm}
\usepackage{amsfonts}
\usepackage{gensymb}
\usepackage{romannum}
\usepackage{changes}
\usepackage{titlesec}
\usepackage{etoolbox} 
\usepackage{lipsum} 
\usepackage[capitalize]{cleveref}

\newenvironment{myindentpar}[1]%
{\begin{list}{}%
         {\setlength{\leftmargin}{#1}}%
         \item[]%
}
{\end{list}}

\begin{document}


\title{Catalogue of $C$-paired spin-momentum locking in antiferromagnetic systems}
\author{Mengli HU}
\thanks{Present Address: Institute for Theoretical Solid State Physics, IFW Dresden, Helmholtzstrasse 20, 01069 Dresden, Germany}
\author{Xingkai CHENG}
\author{Zhenqiao HUANG}
\author{Junwei LIU}
\email{liuj@ust.hk}
\affiliation{Department of Physics, Hong Kong University of Science and Technology, Clear Water Bay, Hong Kong, China}

\date{\today}

\begin{abstract}
Antiferromagnetic materials (AFMs) have been gaining lots of attentions due to its great potential in spintronics devices and the recently discovered novel spin structure in the momentum space, i.e., \emph{C}-paired spin-valley or spin-momentum locking (CSVL/CSML), where spins and valleys/momenta are locked to each other due to the crystal symmetry guaranteeing zero magnetization. Here, we systematically studied CSMLs and proposed a general theory and algorithm using little co-group and coset representatives, which reveals that 12 elementary kinds of CSMLs, determined by the geometric relation of spins and valleys and the essential symmetry guaranteeing zero magnetization, are sufficient to fully represent all possible CSMLs. 
By combining the proposed algorithm and high-throughput first-principles calculations, we predicted 38 
magnetic point groups and identified 142
experimentally verified AFMs that can realize CSML.
Besides predicting new materials, our theory can naturally reveal underlying mechanisms of CSMLs' responses to external fields. As an example, two qualitatively different types of piezomagnetism via occupation imbalance or spin tilting were predicted in RbV$_2$Te$_2$O.
The algorithm and conclusions can be directly extended to the locking between valley/momentum and any other pseudo-vector degree of freedom, e.g., Berry curvature, as exemplified in RbV$_2$Te$_2$O and the new proposed piezo-Hall effect, where a strain can induce a non-zero anomalous Hall conductance. In addition, the proposed concept and methodology can be straightforwardly applied to other symmetry groups, such as spin group.

\end{abstract}

\maketitle
\section{INTRODUCTION}

To access the spin degree of freedom (DF) of an electron, it is usually necessary to couple it with other DFs such as local magnetic moment via Zeeman interaction or angular momentum via spin-orbit coupling (SOC). The former one supplies fertile ground for novel properties and applications of ferromagnetic materials (FMs). The latter one has been the driving force of many intriguing phenomena in condensed matter physics and materials sciences in the past few decades, e.g., spin-valley locking (SVL) in two-dimensional materials \cite{xiao2012coupled}, spin texture in topological materials \cite{TI1,TI2}, Rashba effect {\cite{Rashba_effect2}} 
and Dresselhaus effect \cite{Dresselhaus} in semiconductors, which are also the underlying mechanism or building blocking of many thrilling discoveries such as valley Hall effect \cite{xiao2012coupled,2014VHEMak}, spin Hall effect \cite{SHE,SH_mos22019room,SHE_afm2017strong,SHE_afm2021observation}, giant spin-orbit torque \cite{Spin-torque_SH,soctorque_manchon2019current,soctorque_fukami2016magnetization} and Majorana zero modes \cite{MZM_1,MZM_2,MZM_review}. These phenomena seem to be very different, while besides SOC as their same origin, they do share the other same important feature, i.e, the contrasting spin polarization at time-reversal ($T$-) paired momenta due to the Kramers' theorem, and hence can be uniformly named as $T$-paired spin-valley or spin-momentum locking (SVL/SML). It also means that they will suffer from the same problems that spin can only be manipulated by approaches affecting the $T$-symmetry. Consequently, materials with these properties cannot use spin DF to realize the non-volatile information storage by themselves, which have become the bottlenecks hindering their practical applications.
To overcome these problems, in 2021, SVL was extended to the antiferromagnetic systems (AFMs), where valleys or, in general, momenta with contrasting spin splitting are paired by a crystal ($C$-) symmetry instead of the $T$-symmetry, named $C$-paired SVL (CSVL) or SML (CSML) \cite{CSML}, which can come from the Pomeranchuk-type Fermi-surface instabilities in the spin Channel \cite{spincurrent_zhang} or generally exist in AFMs with spin-spitting band structures \cite{CSML,yuan2020giant,CollinearAFM_splitting}.

In CSML materials, any approach affecting the corresponding $C$-symmetry can manipulate spin to induce unconventional phenomena \cite{CSML,sp_sinova2021altermagnetism,sinova2021altermagnetism2,bai2024altermagnetism, Song2025review}. Typically, a strain or electric field can introduce piezomagnetism (PZM) \cite{CSML,spincurrent_zhang} or noncollinear spin current (both spin-polarized current and pure spin current) in collinear AFMs as theoretically predicted \cite{CSML,spincurrent_zhang,spinsplitter_libor,Naka2019} and experimentally confirmed in MnTe \cite{PZM_MnTe} and RuO$_2$ \cite{ruo2_spincurrent,torque_bai2021observation,torque_karube2}.
Besides these novel properties, the predicted spin-splitting band structures \cite{yuan2020giant,CollinearAFM_splitting,CSML} and CSML \cite{CSML} have been directly observed in angle-resolved photoemission spectroscopy in both collinear RuO$_2$ \cite{ruo2_exp,RuO2_ARPES}, MnTe \cite{MnTe_neutron,MnTe_ARPES} and CrSb \cite{CrSb_crystal,CrSb_APRES,CrSb_topological,CrSb_ARPES2,CrSb_ARPES3,CrSb_ARPES4}, Rb/K intercalated V$_2$Te$_2$O \cite{RbV2Te2O_3} and V$_2$Se$_2$O \cite{KV2Se2O}, and noncollinear MnTe$_2$ \cite{MnTe2_neutron, MnTe2_ARPES}.
Moreover, the electrical readout and deterministic $180^{\rm{o}}$ switching of the N$\rm{\acute{e}}$el order has also been realized in Mi$_5$Si$_3$ \cite{application_mn5si3} and CrSb \cite{crsb_transport} thin films with or without assistant magnetic fields via manipulating the corresponding crystal symmetry\cite{crsb_transport}, which further promotes applications of CSML AFMs in ultra-fast, high-density, and highly stable devices. 

To highlight their uniqueness, collinear CSML AFMs were named as altermagnet in 2022 \cite{sp_sinova2021altermagnetism,sinova2021altermagnetism2}, which has dramatically promoted the related study including all the experiments above and many other works. However, all these interesting phenomena above require neither collinear magnetic order nor the absence of spin-orbit coupling, and clearly altermagnet is only a subset of CSML AFMs. Similar terminology has been extended to the non-collinear spins in real space \cite{AM_noncollinear}, while a thorough CSML theory in momentum space still lacks and is in urgent need in understanding unconventional properties of CSMLs and exploring more CSML materials.



In this work, we developed a general and rigorous theory and algorithm to determine all the locked valleys/momenta with contrasting spin polarization, where the point operations in a magnetic space group (MSG) determine the allowed band splitting, the little co-group of a momentum $\overline{G}^{\bm{k}}$ constrains the spin direction, and the coset representatives $\{g_m\}$ gives rise to all the paired momenta with locked spins. We found that all CSMLs can be classified into three types, collinear, coplanar, or spatial, according to the direction of spins associated with the paired momenta, and each type of CSML can be further categorized into four elementary families characterized by the symmetry guaranteeing zero magnetization in AFMs, and these 12 elementary kinds of CSMLs can serve as the building blocks of all CSMLs.
Based on our theory and algorithm, 38 out of 122 MPGs were predicted to host CSML and all possible symmetry-paired momenta and CSMLs are systematically enumerated, which indeed can be completely classified into the 12 elementary kinds as suggested by qualitative analysis. By combining with high-throughput first-principles calculations, 142 CSML materials were screened out of 1794 experimentally verified AFMs in MAGNDATA \cite{data_gallego2016magndata}. The full list can be found in the supplementary material (SM).

Besides cataloguing all CSMLs and predicting new CSML materials, our theory is able to reveal the fundamental difference between symmetries in determining intrinsic properties of CSML materials, which distinguishes from and supplements the conventional symmetry analysis like Neumann's principle that treats all symmetries equally in the same way and hence can lead to unexpected conclusions. Typically, AFMs in the same MSG but with different CSMLs will have qualitatively different properties that are mainly determined by symmetries in little co-group $\overline{G}^{\bm{k}}$ or the coset representatives $\{g_m\}$ but not other symmetries in MSG, which also corresponds to two qualitatively different underlying mechanisms, respectively. Our theory also reveals the importance of Fermi surface, which is critical for many phenomena but usually absent in the conventional symmetry analysis.
We take Rb intercalated V$_2$Te$_2$O as a representative to show its two different types of PZM via either occupation imbalance determined by $\{g_m\}$ or spin tilting determined by $\overline{G}^{\bm{k}}$, which are not distinguished by conventional PZM theories based on the magnetic moments' rotation in real space but can be understood naturally in our new theory.

It is worth noting that the existence of valley DF is not a requirement for any results or conclusions above but will only add additional ingredients to the system and lead to new phenomena like valley (Hall) current
and the spin can be any pseudo-vector DF. This is further exemplified by the Berry curvature distribution of RbV$_2$Te$_2$O as shown later. Based on the $C$-paired locking between Berry curvature and valleys/momenta, we proposed a new piezo-Hall effect, where a strain can break the $C$-symmetry ensuring the exact Berry curvature-momentum locking to induce a non-zero net Berry curvature and hence a finite anomalous Hall conductance. Without any calculations, we can also conclude that the new proposed piezo-Hall effect can generally exist in any CSML materials once spin-orbital coupling or non-collinear magnetic order is considered and this effect can also come from two kinds of mechanisms determined by symmetries in little co-group $\overline{G}^{\bm{k}}$ or the coset representatives $\{g_m\}$, respectively, as in piezomagnetism.

\begin{figure*}[tb!]
  \includegraphics[width=0.9\textwidth]{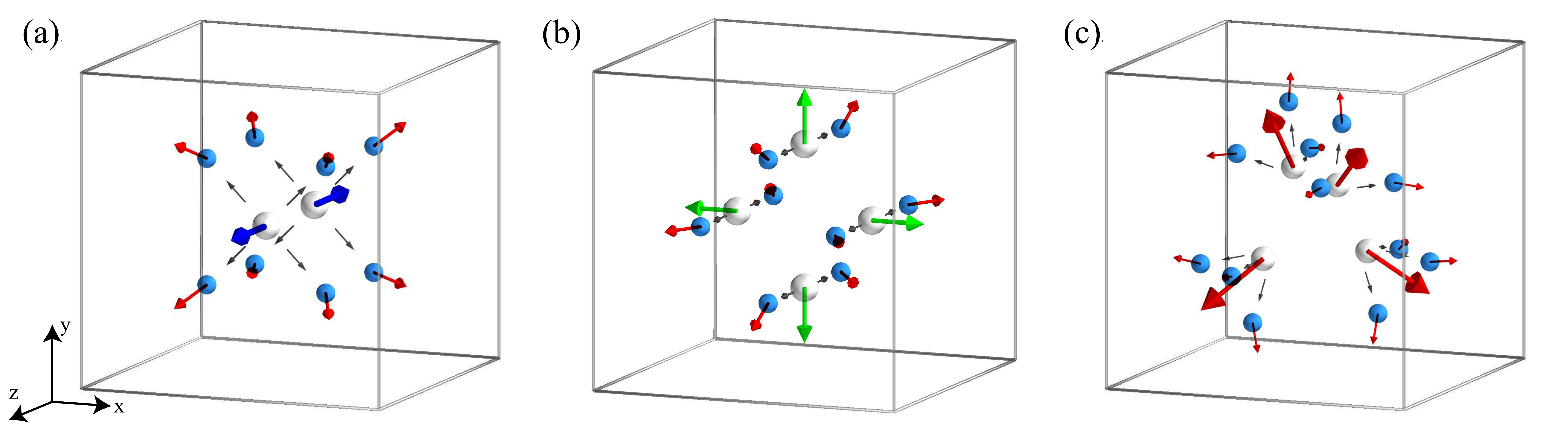}
  \caption{Schematics of three types of CSMLs, collinear (a), coplanar (b) and spatial (c), based on the geometric relation of orientations of spin at the paired valleys.
  White balls indicate the position of valleys in the Brillouin zone (BZ), and the associated arrows represent the corresponding spin orientation of the valley, which can be collinear, coplanar or spatial (non-collinear and non-coplanar). The blue balls and black arrows present the evolution between different types of CSMLs with one valley split into several valleys.
  (a) and (b) show collinear and coplanar CSML in the magnetic point group (MPG) $422$. In (a), spins are along the $z$ direction when the valleys are located at $(0,0,\pm w)$; In (b), spins lie in the $x-y$ plane with valleys at $(\pm u,0,0)$ and $(0,\pm u,0)$. When the valleys deviate from the high-symmetry points, collinear/coplanar CSML transforms to the spatial CSML as shown by the black arrows in (a)/(b). In the spatial CSML (c), the number of coupled valleys in MPG $23$ triples with valleys no longer at $(u, u, u)$ and its equivalent high-symmetry points.}
  \label{fig1}
  \end{figure*}

\section{General theory of CSML}\label{svl theroy}
We first introduce the intuitive classification and criteria for CSMLs. Following the definition, SML can be intuitively classified into three different types, collinear, coplanar and spatial (non-collinear and non-coplanar), according to the relation of spin orientations of equivalent valleys as shown in Fig.~\ref{fig1}, which can also be precisely obtained and identified by our SML algorithm as illustrated later.

Besides the apparent geometric meaning, the classification of three different types has clear physical implications and can directly reveal the qualitative responses of CSML AFMs to external fields. Taking spin current generation \cite{CSML} as an example, the spin polarization of spin currents due to the collinear CSML (Fig.~\ref{fig1}(a)) will be mainly along the $z$ direction, while the spin polarization of spin currents due to the coplanar CSML (Fig.~\ref{fig1}(b)) will be in the $x-y$ plane.
Straightforwardly, different types of CSMLs should be able to realized in the same AFMs depending on the positions of valleys, and the transition between different types of CSMLs can occur as the Fermi level $E_f$ changes. For example, CSML will evolve from a collinear type into a spatial type or further into a coplanar type as depicted by the white and blue balls in Fig.~\ref{fig1} (a) and (b), and consequently the spin polarization of generated spin current will change the direction from the $z$ direction to the $x-y$ plane. It is interesting that a direct transition between a collinear and a coplanar CSML cannot happen but have to go through a spatial CSML.

To rigorously investigate SML and its unconventional properties in AFMs without any prior approximation, one usually needs to consider MSG since it takes into account all possible symmetry constrains and interactions including SOC and effect of non-collinear magnetic orders. However, only point operations (rotation, reflection, and inversion) of space group symmetries, together with the time-reversal operation, can pair momenta and constrain the associated spin orientation (detail derivation in SM I.A). Hence, the simpler MPG instead of more complicated MSG would be enough or even better for studying SML and its corresponding properties. Similarly, if without SOC, the main properties associated with SML will be determined mainly by the spin point group instead of spin space group \cite{sp_liu2022spin,spingroup_1,sp_fangchen2021symmetry,spingroup_2,spingroup_3}. Noted that some MSGs with the same MPGs may have different results since the little co-group of the same momentum might be different for different MSGs even though the MPG is the same, while this will not introduce any new type of CSMLs, i.e. all the possible CSMLs for different MSGs with the same MPG are the same. More details can be found in the SM II. C. 

\begin{figure}[tb]
  \includegraphics[width=0.45\textwidth]{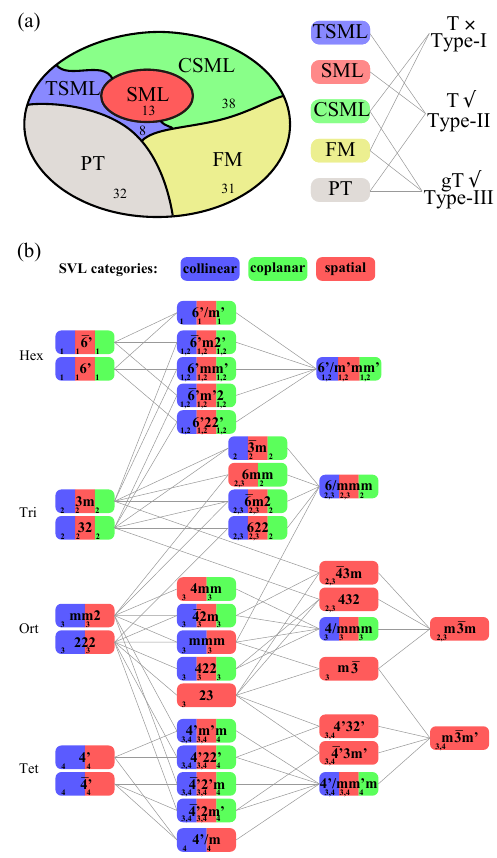}
  \caption{(a) Diagram of MPGs with respect to SML. SML cannot exist in MPGs compatible with FM or $PT$ symmetry. For all 59 MPGs compatible with SML, CSML can exist in 38 MPGs, TSML can exist in 8 MPGs, and CSML and TSML can co-exist in 13 MPGs labeled by SML. The full list of all MPGs can be found in SM 1.D. The connection to the conventional classification \cite{MPG_class1,MPG_class2,MPG_class3} is shown in the right panel. (b) Inheritance of CSMLs among 38 CSML MPGs. Grey solid line connects a MPG and its subgroups. The background color denotes three types of CSML and the numbers in the bottom left corner of each type represent the four elementary families. Hermann-Mauguin notation is used.}
  \label{fig2}
\end{figure}

The primary criterion for a material to exhibit CSML is the existence of spin-splitting band structures, which can naturally occur in magnetic materials due to exchange interactions between electrons and localized magnetic moments once parity (inversion) times time-reversal ($PT$) symmetry is broken \cite{sp_hayami2020,yuan2020giant, liu2023universal}.
Secondly, we focus on CSML and hence exclude MPGs that are compatible with TSML since the main properties will be determined by TSML instead of CSML once it exists. 
TSML magnetic materials will be very similar to nonmagnetic (NM) materials in many aspects because there is an effective time-reversal symmetry $\tilde{T}^2=-1$ even the pure real time reversal symmetry is broken due to the magnetic structure. For example, in AFMs with checkerboard and N$\rm{\acute{e}}$el order, $T$ times fractional translation $\tau$ symmetry $\tilde{T}=T\tau$ can be preserved.
The final criterion is that at least one symmetry except $PT$ symmetry must exist to ensure valleys and spins strictly locked to each other, which, will be shown later, is actually the symmetry that guarantees the exactly zero net magnetization in AFMs and hence can be naturally realized in AFMs.

Based on these three criteria above, all the MPGs can be classified into five classes depending on the existence of spin-splitting band structures and different types of SML as shown in Fig.~\ref{fig2} (a). 
The first class includes all the 32 MPGs with $PT$ symmetry, and all the bands in the whole BZ must be spin degenerate. The other four classes will allow spin-splitting bands at a generic momentum. Among them, the second class is compatible with a non-zero magnetization and hence can realize FM, weak FM (canted AFM) or ferrimagnetism.
Noted that the total magnetization in these MPGs could be zero due to the magnetic anisotropy energy, and hence the rigorous SML can be also realized. However, it would be materials dependent and the spin-dependent properties would possess both FM and AFM nature. Many experimentally realized altermagnets are belong to this class due to a vanishingly small but finite magnetization and hence can realize anomalous Hall effect that should be absent in perfect altermagnets or CSML materials. A simple example is a two-dimension system that is compatible with MPG or MSG allowing out-of-plane net magnetization can be a perfect AFM if magnetic anisotropy energy favors the in-plane magnetic order and enforces an exact zero total magnetization. As a real material example, CoNb$_3$S$_6$ of MPG 32' is an AFM in experiment although MPG 32' is compatible with FM and allows a finite magnetization.

Based on the symmetries ensuring the zero magnetization and pairing equivalent momenta, the left MPGs can be classified into another three classes: 38 CSML MPGs as listed in Fig.~\ref{fig2} (b) are due to crystal symmetries, 8 TSML MPGs are due to (effective) time reversal symmetry, and the left 13 MPGs are due to both crystal symmetries and time-reversal symmetry. It is worth noting that except 38 CSML MPGs that must be realized in AFMs, the 21 MPGs hosting TSML and the 32 PT MPGs can exist in both AFMs and NMs. In this aspect, AFMs and NMs are indistinguishable in momentum space for 21 MPGs hosting TSML and 32 PT MPGs, and many of their properties are also qualitatively the same, which contrasts conventional understandings. As a result, one has to rely on the information or properties in real space and use MSGs to distinguish these AFMs and NMs.

\subsection{SML algorithm and its application}\label{SML algorithm}
Now we introduce the rigorous SML algorithm that can determine all the paired momenta and the orientation of associated spins. We found that two ingredients, the little co-group of momentum $\overline{G}^{\bm{k}}$ and the coset representatives $\{g_m\}$, can fully capture and determine all the features of any SML with the former uniquely constraining the spin's orientation of the momentum $\bm{k}$ and the latter fully determining all the paired momenta and the associate spins' relation, respectively.
The full SML algorithm consists of three steps:
\begin{myindentpar}{12pt}
(1) Get the MPG of a given AFM, labelled as $\overline{G}$;

(2) For a momentum $\bm{k}$, find its little co-group $\overline{G}^{\bm{k}}$ by checking if a element $g$ in $\overline{G}$ satisfies $\bm{k} = g\bm{k} + \bm{K}$, where $\bm{K}$ is a reciprocal vector.
CSML can only exist for momenta where spin splitting or spin polarization $\bm{s}$ is allowed, and the necessary and sufficient condition is that $\overline{G}^{\bm{k}}$ is compatible with FM, i.e. belongs to 31 FM MPGs as shown in Fig.~\ref{fig2} (a) and listed in SM 1.D, otherwise such momenta will be spin-degenerate and cannot realize SML;

(3) All the paired momenta $\{\bm{k}^*\}$ and associate spins $\{\bm{s}^*\}$ can be obtained by $\{\bm{k}^*;\bm{s}^*\} = \{g_m \bm{k};g_m \bm{s}\}$, where $g_m$ is the representative of coset $g_m\overline{G}^{\bm{k}}$ and can be obtained via the coset decomposition $\overline{G}=\sum_m g_m\overline{G}^{\bm{k}}$. 
The matrix presentation of all $g_m$ can be found in SM I.C.
\end{myindentpar}

We applied this algorithm for all MSGs to enumerate all possible CSMLs, and found that all CSMLs indeed belong to collinear ($L$), coplanar ($P$), or spatial ($S$) as shown in Fig.~\ref{fig1}, which confirms the intuitive analysis above.
The CSML type can be rigorously determined based on $\overline{G}^{\bm{k}}$ without knowing other details as follows:
\begin{align}
  & \forall \bm{k}, \exists (-)N_{\bm{\alpha}} \in \overline{G}^{\bm{k}} \leftrightarrow L, \label{eq:eq1}\tag{C.1}\\
  & \forall \bm{k}, \exists (-)N'_{\bm{\alpha}} \in \overline{G}^{\bm{k}} \rightarrow P,\label{eq:eq2}\tag{C.2}\\
  & \forall \bm{k}, \exists (-)N_{\bm{\alpha_k}} \in \overline{G}^{\bm{k}}, {\bm{\alpha_k}} \perp {\bm{\alpha}} \rightarrow P, \label{eq:eq3} \tag{C.3}
\end{align}
where $\bm{k} \in \{\bm{k}^*\}$, $-$ is the inversion operation, $N_{\bm{\alpha}}$ represents the $N-$fold rotation operation ($N>1$) about $\bm{\alpha}$ axis and $N'$ represents $N-$fold rotation times time-reversal symmetry. 
$\leftrightarrow$ denotes the necessary and sufficient and $\rightarrow$ denotes the sufficient condition, respectively.

We take MPG $\overline{G}=422$ (in Hermann-Mauguin notation) and momentum $\bm{k} = (0,0,w)$ as an explicit example as shown in Fig.~\ref{fig1} (a).
Based on the generators ($4_{001}$ and $2_{100}$) of $422$, the little co-group of momentum $\bm{k} = (0,0,w)$ can be easily obtained as $\overline{G}^{\bm{k}}=\{1,4^+_{001},2_{001},4^-_{001}\}=4$.
Clearly, $4$ is among 31 FM MPGs listed in SM 1.D and belongs to FM, thus CSML can exist. 
Then via the coset decomposition $\overline{G} = \overline{G}^{\bm{k}} + 2_{100}\overline{G}^{\bm{k}}$, the coset representatives can be obtained as $\{g_m\} = \{1,2_{100}\}$, which gives all the paired valleys and spins as $\{\bm{k}^*;\bm{s}^*\} = \{ (0,0,\pm w); (0,0,\pm m_z) \}$, belonging to collinear CSML.
The same conclusion can also be drawn by $\overline{G}^{\bm{k}}$ since the condition \ref{eq:eq1} is satisfied as $\forall \bm{k}, \ 4_{001}\in \overline{G}^{\bm{k}}$. Now we can obtain a counterintuitive conclusion that any magnetic materials with $\overline{G}=422$ and valleys/momentum located at $\bm{k} = (0,0,w)$ belong to collinear CSML even if their magnetic order in real space is noncollinear or noncoplanar, e.g. coplanar AFM Ho$_2$Ge$_2$O$_7$.
The same procedure was conducted for all the valleys/momenta of 38 CSML MPGs, and the results are summarized in Fig.~\ref{fig2}.
Counterintuitively, a cubic crystal can only allow the spatial CSML although it possesses the highest symmetry group.
The reason is that cubic crystal hosts three-fold rotational symmetries along body diagonals which violates all the conditions allowing collinear (\ref{eq:eq1}) or coplanar CSML (\ref{eq:eq2} and \ref{eq:eq3}).  

Besides enumerating all the CSMLs, our algorithm can rigorously capture the transition between different types of CSMLs, which can be described by another coset decomposition and is fully predictable.
Taking the collinear CSML in Fig.~\ref{fig1}(a) as an example, a momenta at $\bm{k} = (0,0,w)$ splits into several general momenta, $\bm{k} = (0,0,w) \rightarrow \bm{k'} = (u,v,w)$. Such a splitting can be described as 
$\overline{G}^{\bm{k}} = \overline{G}^{\bm{k'}} + 4^+_{001}\overline{G}^{\bm{k'}} + 2_{001}\overline{G}^{\bm{k'}} + 4^-_{001}\overline{G}^{\bm{k'}}$, which tells that each momentum splits into four momenta paired by the coset representatives as $\{\bm{k'}^*;\bm{s'}^*\} = \{\bm{k'},4^+_{001}\bm{k'},2_{001}\bm{k'},4^-_{001}\bm{k'};\bm{s'},4^+_{001}\bm{s'},2_{001}\bm{s'},4^-_{001}\bm{s'} \}$ and the CSML transits from the collinear type into the spatial type. 
As a thorough summary of all the CSMLs of 38 MPGs in Fig.~\ref{fig2}, one can predict and design CSMLs and the transition accordingly.

\subsection{Elementary family of CSML}
Although all CSMLs can be completely enumerated by our SML algorithm above, the connection between CSMLs of different MPGs is still unclear. Typically, some types of CSMLs only exist in a subgroup but \emph{not} in its parent group and vise versa as shown in Fig.~\ref{fig2}, which stems from a simple fact that the little co-group of a subgroup is not necessarily a subgroup of the little co-group of its parent group. Such an inconsistency also means that the classification merely using three different types cannot fully uncover the relation of all CSMLs. 

To address this inadequacy, we carefully examined all the CSMLs enumerated by our SML algorithm above and found that 
(1) MPGs that can host CSMLs have at least one subgroup as $\pm 6'$, $32/3m$, $mm2/222$, and $\pm 4'$, which are also four families of the smallest groups compatible with AFMs to guarantee the zero net magnetization, and hence we define these MPGs as four families of elementary CSML MPGs; 
(2) The four families of elementary CSML MPGs correspond to four different crystal families, hexagonal, trigonal, orthorhombic, and tetragonal crystals, respectively; 
(3) CSMLs that can be realized in these four families of elementary CSML MPGs come from the simplest $\overline{G}^{\bm{k}}$ and $\{g_m \}$ and have the smallest number of paired valleys/momenta; 
(4) For any collinear and coplanar CSML, the symmetry axes in $\{g_m \}$ and $\overline{G}^{\bm{k}}$ must be orthogonal or parallel, and symmetries in $\{g_m \}$ and $\overline{G}^{\bm{k}}$ can be broken separately or simultaneously in orthogonal CSMLs but must be broken simultaneously in parallel CSMLs. Meanwhile, both orthogonal and parallel CSMLs can exist in the same system whose $\{g_m \}$ and $\overline{G}^{\bm{k}}$ can be broken separately. 
(5) Most importantly, a complicated CSML with more paired valleys/momenta that can be realized in other MPGs can always be taken as multiple copies of the simplest CSMLs realized in the elementary CSML MPGs. Based on these discoveries, we propose four elementary families of CSMLs as a new dimension and feature to describe all the CSMLs.

\begin{figure}[tb]
	\centering
	\includegraphics[width=0.45\textwidth]{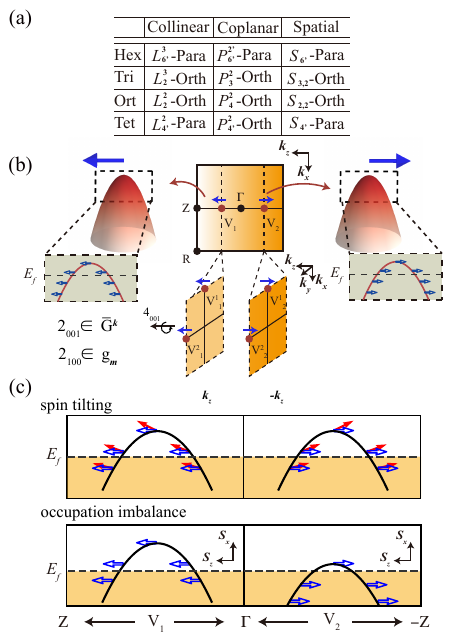}
	\caption{(a) Table of 12 elementary kinds of CSMLs. Four elementary families of CSMLs are labeled by the abbreviation of hexagonal, trigonal, orthorhombic, and tetragonal, and different types of CSMLs with each elementary family are labeled by $L/P/S^{N_1(')}_{N_2(')}$, where $N_{1/2}(')$ labels the $N_{1/2}-$ fold rotational symmetry in the $\overline{G}^{\bm{k}}$ and $\{g_m\}$, respectively. 
		(b) Illustration of Ort-$L_2^2$-Orth kind CSML in MPG $422$. The band structures around the valleys $V_{1/2}$ located at $(0,0,\pm w)$ and spins along $\bm{z}$ are presented. In the lower panel, the same kind CSML located at $(0,\frac{1}{2},\pm w)$ and $(\frac{1}{2},0,\pm w)$ are shown. Both CSML cases are paired by symmetry $2_{100}/2_{010}$ for their same CSML kind and symmetry $4_{001}$ merely doubles the number of paired valleys by generating another copy ($V^2$) of the elementary CSML ($V^1$) but does not introduce any new property. 
		(c) Two qualitatively different mechanisms generating PZM in Ort-$L_2^2$-Orth: spin tilting presented by red filled arrow (top panel) and occupation imbalance reflected by the relative different Fermi levels of the valleys (bottom panel). }
	\label{fig3}
\end{figure}

Together with three different types proposed above, 12 elementary kinds of CSMLs listed in the table in Fig.~\ref{fig3} (a) can completely characterize and serve as building blocks for all the CSMLs as also indicated by the subscript numbers in the left bottom corner of each type of CSMLs in 38 CSML MPGs in Fig.~\ref{fig2}. They can be concretely labeled as $F-T^{N_1(')}_{N_2(')}-R$, where $F$=Hex, Tri, Ort, or Tet stands for the hexagonal, trigonal, orthorhombic or tetragonal family, and $T=L, P$ or $S$ means collinear, coplanar or spatial type, the superscript $N_1(')$ and subscript $N_2(')$ represent generators in $\overline{G}^{\bm{k}}$ and $\{g_m \}$, and $R=$ Orth or Para means orthogonal or parallel between rotation axes of $N_1(')$ and $N_2(')$. Noted that for a more compact notation, the suffix $-R$ can be removed since the 12 elementary kinds can only belong to either -Orth or -Para but not both. 

Now it is clear that it is the elementary family instead of the geometric relation of spins' orientation inherited between different CSML MPGs as shown in Fig.~\ref{fig2}. For example, MPG $432$ can only realize the spatial-type CSMLs, while it can realize both trigonal and orthorhombic CSMLs since its subgroups contain $32$ and $222$. On the contrary, although $4/mmm$ can realize all collinear, coplanar and spatial types of CSMLs, it can only realize orthorhombic CSML because its subgroups only contain $mm2$/$222$ but not other elementary CSML MPGs.
This inheritance relation between elementary CSMLs and more complicated CSMLs can be also rigorously described by another coset decomposition as $\overline{G}=\sum_r g_r\overline{G}^E$, where $\overline{G}^E$ is the elementary CSML MPG and $\{g_r\}$ are coset representative. The symmetries in $\{g_r\}$ will only generate additional copies of the elementary CSMLs determined by $\overline{G}^E$ but not introduce too much new physics.
As exemplified by the collinear CSMLs in $422$ as shown in Fig.~\ref{fig3} (b), a complicated collinear CSML with four different valleys/momenta at $k=(0,\frac{1}{2},\pm w)$ and $(\frac{1}{2},0,\pm w)$ are paired by $2_{100}/2_{010}$ and $4_{001}$, while only $2_{100}/2_{010}$ can flip the direction of spins and enforce the opposite spin polarization for valleys/momenta at $k=(0,\frac{1}{2}, \pm w)$ or $(\frac{1}{2}, 0, \pm w)$ and the $4_{001}$ symmetry merely generates another copy of the elementary CSML paired by $2_{100}/2_{010}$, i.e. Ort-$L^2_2$-Orth CSML in MPG $222$, which can be rigorously obtained by the coset decomposition $422 = 222 + 4_{001} 222$. Therefore, the collinear CSML in $422$ will have the same property as the elementary Ort-$L^2_2$-Orth CSML in MPG $222$. Breaking the symmetry $4_{001}$ will simply reduce the number of paired valleys/momenta from four to two and only the symmetry $2_{100}/2_{010}$ matters for the unconventional properties arisen from CSMLs.

More importantly, without any complicated symmetry analysis and calculation, the elementary CSML has already been able to capture the main properties of AFMs arisen from CSMLs. To elucidate this more clearly, we will take PZM as an example.
In the example of collinear CSMLs in $422$ and $222$ that are Ort$-L^{N_1(')}_{N_2(')}-$Orth CSML, they belong to orthogonal CSMLs, which means the symmetries in $\overline{G}^{\bm{k}}$ and $\{g_m\}$ are orthogonal to each other and can be broken separately to realize two qualitatively different types of PZM. As illustrated in the top panel of Fig.~\ref{fig3} (c), breaking symmetry in $\overline{G}^{\bm{k}}$ will induce spin tilting for all the valleys/momenta to induce uncompensated net magnetization, although different valleys/momenta are still equivalent due to $\{g_m\}$, and the direction of net magnetization will depend on $\{g_m\}$ and hence be in $x-y$ plane for the case in Fig.~\ref{fig3} (c). This is like the conventional PZM due to local magnetic moments' nonequivalent rotation under strain \cite{PZM_1994} in real space, but it occurs in momentum space for spin of itinerant electrons associated with different valleys/momenta.
On the other hand, as breaking symmetry in $\{g_m\}$, the strain will break the equivalence of different valleys/momenta to induce different energy shifts and hence lead to unequal occupation of spin-up and spin-down electrons, and the net magnetization should be along $z$ direction since spin's orientation remains unchanged as shown in the bottom panel of Fig.~\ref{fig3} (c). We refer these two microscopic mechanisms as spin tilting and occupation imbalance with the former driving AFM into weak FM (canted AFM) and the latter driving AFM into ferrimagnetism from the aspect of the real space. Notably, these two types of PZMs may coexist for certain strains, and both are qualitatively different from the conventional theory since the non-zero net magnetization due to either spin tilting or occupation imbalance are from itinerant electrons instead of local magnetic moments \cite{PZM_1994} and hence can only happen in conducting AFMs.

In a short summary of our theory, with three different types and four elementary families as two independent features and dimensions, all CSMLs can be fully characterized and classified into 12 elementary kinds, and the main properties of CSML AFMs will be determined by the elementary CSMLs realized in these materials instead of other factors or material details.

\begin{figure*}[htb!]
  \centering
    \includegraphics[width=0.8\textwidth]{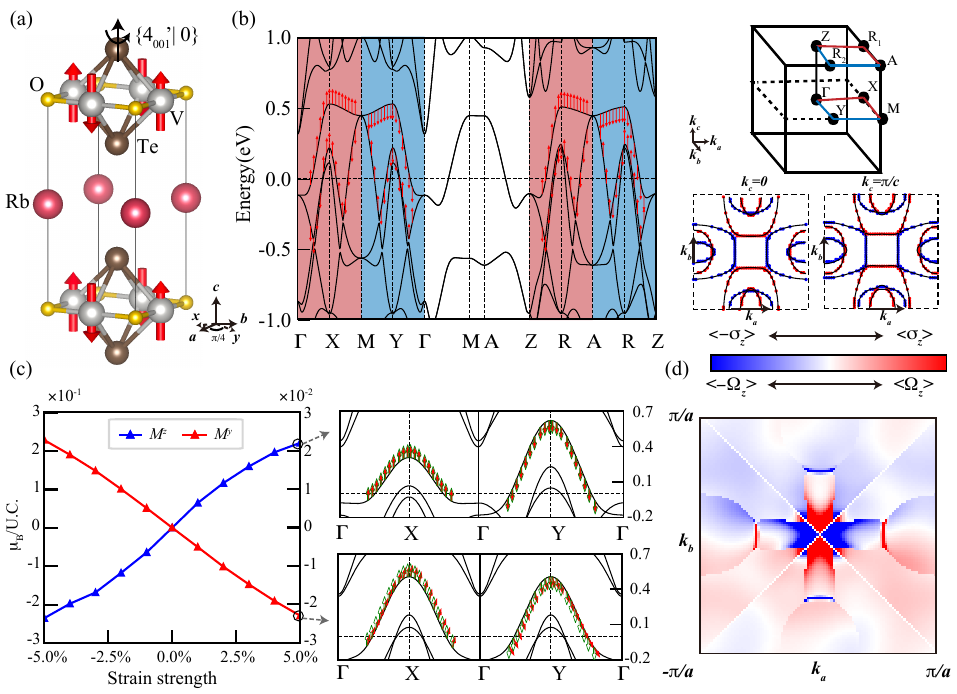}
      \caption{\label{fig4} 
  A representative CSML material Rb intercalated V$_2$Te$_2$O. (a) Crystal and magnetic structure of RbV$_2$Te$_2$O, where the magnetic moments are labeled by red arrows. The Cartesian coordinate is anticlockwise rotated by $\pi/4$ of the unit vector along $\bm{c}$ direction. (b) Band structure of RbV$_2$Te$_2$O along high-symmetry lines, where the region of coupled valleys are labelled by shadowed red and blue. The spin orientations of bands across Fermi level are denoted. The spin polarized Fermi surfaces are plotted in the lower right panel at $k_c=0$ (left) and $k_c=\frac{\pi}{\bm{c}}$ (right), respectively. The 1$^{st}$ BZ is shown in the upper right panel with high-symmetry points and corresponding spin polarization denoted. (c) Two types of PZM generated due to different mechanisms under two different kinds of strain $M^y = \Lambda^y_{xz} \sigma_{xz}$ and $M^z = \Lambda^z_{xx} \sigma_{xx}$. The strained valley band structures under $+5\%$ strain strength are shown in the right panels. The spin orientations without and with strain are labeled by solid red and hollow green arrows, respectively.
(d) The Berry curvature($\Omega$) distribution in the 1$^{st}$ BZ with $k_c = 0$. The color-map ranging from blue to red indicates the strength of $\Omega_z$.}
\end{figure*}

\section{Application and material realization}\label{3d materials}
With the complete CSML theory and algorithm above, we employed the high-throughput first-principles calculations to predict CSML material candidates from the magnetic database MAGNDATA \cite{data_gallego2016magndata} and identified 142 CSML candidates with the experimentally confirmed desired magnetic structures. The full list of 142 CSML materials and their band structures along high-symmetry lines can be found in SM III.

In the following discussion, we choose Rb intercalated V$_2$Te$_2$O in our database as a representative since it is one of the mostly studied AFMs now and focus on its PZM which has not been fully studied.
As shown in Fig.~\ref{fig4}(a), RbV$_2$Te$_2$O belongs to the tetragonal crystal \cite{RbV2Te2O_1,RbV2Te2O_2,RbV2Te2O_3} and is a collinear AFM with N\'{e}el temperature higher than 300 K \cite{RbV2Te2O_3}.
The local magnetic moments of V atoms are along the $z$ direction, and two  magnetic sublattices formed by different V atoms are connected by magnetic symmetry $\{4'_{001} | 0\}$ and two diagonal mirror symmetry, which guarantee the perfect compensation and zero magnetization. The adjacent magnetic layers share the same magnetic order, forming the $C-$type AFM and its MPG is $\overline{G} = 4'/mm'm$.

As theoretically predicted and experimentally verified \cite{CSML,RbV2Te2O_3}, RbV$_2$Te$_2$O is metallic with spin-splitting Fermi surfaces along $(\pm u,0,v)$ and $(0, \pm u,v)$ as shown in Fig.~\ref{fig4} (b). RbV$_2$Te$_2$O is a Van der Waals layered materials and the dispersion along $z$ direction is almost unnoticeable, and hence we will focus on $k_z=0$ or $k_z=0.5$ planes.
The little co-group at $k = (\pm u,0,0/0.5)$ and $(0, \pm u,0/0.5)$ is $\overline{G}^{\bm{k}} = m'm2' $, and clearly $\forall \bm{k}, m \in \overline{G}^{\bm{k}}$, which satisfies the collinear condition \ref{eq:eq1}. Together with Fig.~\ref{fig2}, without any calculation of spin polarization, one can recognize that RbV$_2$Te$_2$O can realize the elementary CSML Tet-$L^2_{4'}$-Para and Ort-$L^2_{2}$-Orth, which is confirmed by the spin textures obtained by the first-principles calculations as denoted by red arrows in Fig.~\ref{fig4} (b). As a result, RbV$_2$Te$_2$O can realize two different types of PZMs with spin tilting or occupation imbalance under different strains.
To confirm it, two different types of strains $\sigma_{xx}$ and $\sigma_{xz}$ are considered, which can separately break either $\{g_m\}$ or $\overline{G}^{\bm{k}}$ in Ort-$L^2_{2}$-Orth and correspond to occupation imbalance or spin tilting, respectively. The results obtained by first-principles calculations with SOC considered are shown in Fig.~\ref{fig4} (c). Both types of strains induce a net magnetization with the direction of magnetization and the microscopic mechanisms shown in the right panel of Fig.~\ref{fig4} (c). Under the uniaxial stain $\sigma_{xx}$ that only breaks the symmetry connecting different valley $\{g_m\}$ but maintain symmetries in little co-group $\overline{G}^{\bm{k}}$, the spins are still along $z$ direction but the occupation of spin-up and spin-down electrons are clearly imbalanced and hence the magnetization is strictly along $z$ direction, and the strained RbV$_2$Te$_2$O will behave like a ferrimagnet. On the contrary, a shear strain $\sigma_{xz}$ can break
$\overline{G}^{\bm{k}}$ but reserve $\{g_m\}$, and hence the spins around valleys all tilt from $\bm{z}$ direction to $x-y$ plane, although two valleys are still equivalent under a shear strain $\sigma_{xz}$. As a result, the total magnetization will be in-plane. In this case, the strained RbV$_2$Te$_2$O is more like a weak FM or tilted AFM. 
These uniaxial ($\sigma_{xx}$) and shear ($\sigma_{xz}$) strain induced PZM are consistent with previous study of linear symmetry-imposed PZM tensor\cite{MPG_class2,Bilbao_tensor}, where the nonzero elements are $\Lambda^a_{ac} = \Lambda^a_{ca} = -\Lambda^b_{bc}= -\Lambda^b_{cb}, \ \Lambda^c_{aa} = -\Lambda^c_{bb}$, and $M^i = \Lambda^i_{jk}\sigma_{jk}$. Similar results and conclusions can also be explicitly demonstrated in KV$_2$Se$_2$O \cite{KV2Se2O} and RuO$_2$ as shown in SM IV. These results clearly confirm two different microscopic mechanisms of PZMs revealed by our general theory of CSMLs as explained in previous sessions and hence demonstrate the validity of our general theory and algorithm.

Since any pseudo-vector DF obeys the same transformation as spin under point symmetry operations, our theory and algorithm above can be directly applied to the locking between valley/momentum and any pseudo-vector DF. Here, we use Berry curvature $\bm{\Omega}$ of RbV$_2$Te$_2$O as an example. This $\bm{\Omega}$-momentum locking phenomenon was first proposed in the TSVL 2H-MoS$_2$ \cite{xiao2012coupled} and is the origin of valley Hall effect \cite{2014VHEMak}. Based on our CSML results, the $\bm{\Omega}$-momentum locking should be the same elementary kind, i.e., Tet-$L^2_{4'}$-Para and Ort-$L^2_{2}$-Orth. 
In addition, as $\overline{G} = \{\overline{G}^E,-1\overline{G}^E,2_{110}\overline{G}^E,m_{110}\overline{G}^E\}$ and $\overline{G}^E = 4'$, there are four copies of $L^{2}_{4'}$ after the decomposition with respect to the elementary kind of CSML.
The distribution of $\bm{\Omega}(k_x,k_y,0)$ obtained by first-principles calculations is shown in Fig.~\ref{fig4}(d), which indeed follows the prediction from our algorithm. These fan-shaped regions with same magnitude and alternating signs of $\Omega_z$ are four valleys belonging to $L^{2}_{4'}/L^2_{2}$ class, which will generate valley or spin Hall effect that are expected to reveal richer phenomena or properties of CSML materials.
Importantly, as long as the AFM symmetry is strictly preserved, the total ${\bm{\Omega}}$ must be zero and the anomalous Hall effect is forbidden. However, same as the PZM discussed above, a symmetry breaking induced by external strains will lead to non-zero net ${\bm{\Omega}}$ via either imbalanced occupation or $\bm{\Omega}$ tilting and hence the corresponding anomalous Hall effect, which can be referred to as piezo-Hall effect.
Besides, other approaches breaking the same crystal symmetries such as intrinsic distortions or finite magnetization due to Dzyaloshinsky-Moriya interaction can also generate net ${\bm{\Omega}}$ and the anomalous Hall effect.

\section{Summary and perspective}
The new terminology, altermagnet, has attracted extensive attention and greatly promoted the study of spin-splitting AFMs or more broad unconventional magnetism. However, there are many inaccurate usage of altermagnet, which did not distinguish from canted AFMs and ferrimagnets, and hence leads to many inaccurate conclusions. For example, whether anomalous Hall effect can exist in altermagnets or not \cite{qihang2025,Song2025review}. Compared to traditional study of magnetic materials focusing on real space, our theory shows that the key advances lie in the understanding of the spin structure in momentum space, e.g., CSML, where symmetries involving crystal operations enforce a perfect locking between spin and momentum. It is worth emphasizing that CMSL is fundamentally and intrinsically different from momentum-dependent spin splitting. In fact, spin splitting is always momentum dependent in any kind of materials as long as bands are not spin-degenerate as in FMs, ferrimagnets and even non-magnetic materials with spin-orbit coupling or under external magnetic fields, and even the sign of such spin splitting can also be opposite at different momenta in these kinds of materials with suitable orbitals and interaction strengths. On the contrary, CSML can only exist in a perfect AFM or altermagnet but not any other materials. CSML ensures exact compensation of spin polarizations in momentum space and hence the exactly zero net magnetization but cannot lead to the exact cancellation of responses of spin-up and spin-down electrons to external fields as strain and electric fields, and hence CSML systems can have many FM-like responses even though the total magnetization is indeed zero, which fundamentally distinguish from responses in ferrimagnets and canted AFMs. The differences among these concepts seem small but critical in the rigorous discussions of the novel properties of altermagnets.

With a focus on CSML, our work provides a unified and rigorous framework and algorithm to understand unconventional magnetism, which can be generally applied in all unconventional magnetism independent of whether the magnetic order is collinear and whether spin-orbit coupling is considered and can be extended to all spin-like pseudo-vector DFs. Although the number of all possible CSMLs seems to be very large or even infinite, we demonstrated that there are only 12 elementary kinds of CSMLs that can be rigorously defined by the geometric relation of spins and the essential symmetry guaranteeing the zero net magnetization. These 12 elementary kinds of CSMLs can capture all the new physics due to CSMLs and can serve as building blocks for all CSMLs. Our theory can easily and rigorously clarify many controversial conclusions. For example, we can conclude that a perfect altermagnet or CSML system intrinsically forbids the anomalous Hall effect since Berry curvature follows the same rules as magnetization under symmetry operations, while the anomalous Hall effect can be induced by breaking the symmetry that leads to CSML via external strain or magnetic fields or internal Dzyaloshinskii–Moriya interaction, which effectively introduce a net magnetization and hence drive a perfect CSML system or altermagnet to a canted AFM or a ferrimagnet even though the net magnetization might be vanishingly small and does not contribute to the Berry curvature. Moreover, based on our theory, many new phenomena and properties of CSML materials can be predicted and understood, such as  piezo-Hall effect, valley Hall effect, piezo-magnetoelectric effect, and multiferroic effect, which can be experimentally verified and have potential applications in spintronics and valleytronics. The predicted 142 CSML materials as listed in SM and their properties will provide experimentalist more material platforms and may become a new research frontier in the field of magnetism, materials sciences, and condensed matter physics.

\section{Acknowledgement}
This work is supported by National Key R\&D Program of China (Grants No.2021YFA1401500) and the Hong Kong Research Grants Council (16303821, 16306722 and 16304523).


\begin{thebibliography}{63}%
\makeatletter
\providecommand \@ifxundefined [1]{%
 \@ifx{#1\undefined}
}%
\providecommand \@ifnum [1]{%
 \ifnum #1\expandafter \@firstoftwo
 \else \expandafter \@secondoftwo
 \fi
}%
\providecommand \@ifx [1]{%
 \ifx #1\expandafter \@firstoftwo
 \else \expandafter \@secondoftwo
 \fi
}%
\providecommand \natexlab [1]{#1}%
\providecommand \enquote  [1]{``#1''}%
\providecommand \bibnamefont  [1]{#1}%
\providecommand \bibfnamefont [1]{#1}%
\providecommand \citenamefont [1]{#1}%
\providecommand \href@noop [0]{\@secondoftwo}%
\providecommand \href [0]{\begingroup \@sanitize@url \@href}%
\providecommand \@href[1]{\@@startlink{#1}\@@href}%
\providecommand \@@href[1]{\endgroup#1\@@endlink}%
\providecommand \@sanitize@url [0]{\catcode `\\12\catcode `\$12\catcode `\&12\catcode `\#12\catcode `\^12\catcode `\_12\catcode `\%12\relax}%
\providecommand \@@startlink[1]{}%
\providecommand \@@endlink[0]{}%
\providecommand \url  [0]{\begingroup\@sanitize@url \@url }%
\providecommand \@url [1]{\endgroup\@href {#1}{\urlprefix }}%
\providecommand \urlprefix  [0]{URL }%
\providecommand \Eprint [0]{\href }%
\providecommand \doibase [0]{https://doi.org/}%
\providecommand \selectlanguage [0]{\@gobble}%
\providecommand \bibinfo  [0]{\@secondoftwo}%
\providecommand \bibfield  [0]{\@secondoftwo}%
\providecommand \translation [1]{[#1]}%
\providecommand \BibitemOpen [0]{}%
\providecommand \bibitemStop [0]{}%
\providecommand \bibitemNoStop [0]{.\EOS\space}%
\providecommand \EOS [0]{\spacefactor3000\relax}%
\providecommand \BibitemShut  [1]{\csname bibitem#1\endcsname}%
\let\auto@bib@innerbib\@empty
\bibitem [{\citenamefont {Xiao}\ \emph {et~al.}(2012)\citenamefont {Xiao}, \citenamefont {Liu}, \citenamefont {Feng}, \citenamefont {Xu},\ and\ \citenamefont {Yao}}]{xiao2012coupled}%
  \BibitemOpen
  \bibfield  {author} {\bibinfo {author} {\bibfnamefont {D.}~\bibnamefont {Xiao}}, \bibinfo {author} {\bibfnamefont {G.-B.}\ \bibnamefont {Liu}}, \bibinfo {author} {\bibfnamefont {W.}~\bibnamefont {Feng}}, \bibinfo {author} {\bibfnamefont {X.}~\bibnamefont {Xu}},\ and\ \bibinfo {author} {\bibfnamefont {W.}~\bibnamefont {Yao}},\ }\bibfield  {title} {\bibinfo {title} {{Coupled Spin and Valley Physics in Monolayers of ${\mathrm{MoS}}_{2}$ and Other Group-VI Dichalcogenides}},\ }\href {https://doi.org/10.1103/PhysRevLett.108.196802} {\bibfield  {journal} {\bibinfo  {journal} {Phys. Rev. Lett.}\ }\textbf {\bibinfo {volume} {108}},\ \bibinfo {pages} {196802} (\bibinfo {year} {2012})}\BibitemShut {NoStop}%
\bibitem [{\citenamefont {Hasan}\ and\ \citenamefont {Kane}(2010)}]{TI1}%
  \BibitemOpen
  \bibfield  {author} {\bibinfo {author} {\bibfnamefont {M.~Z.}\ \bibnamefont {Hasan}}\ and\ \bibinfo {author} {\bibfnamefont {C.~L.}\ \bibnamefont {Kane}},\ }\bibfield  {title} {\bibinfo {title} {Colloquium: Topological insulators},\ }\href {https://doi.org/10.1103/RevModPhys.82.3045} {\bibfield  {journal} {\bibinfo  {journal} {Rev. Mod. Phys.}\ }\textbf {\bibinfo {volume} {82}},\ \bibinfo {pages} {3045} (\bibinfo {year} {2010})}\BibitemShut {NoStop}%
\bibitem [{\citenamefont {Qi}\ and\ \citenamefont {Zhang}(2011)}]{TI2}%
  \BibitemOpen
  \bibfield  {author} {\bibinfo {author} {\bibfnamefont {X.-L.}\ \bibnamefont {Qi}}\ and\ \bibinfo {author} {\bibfnamefont {S.-C.}\ \bibnamefont {Zhang}},\ }\bibfield  {title} {\bibinfo {title} {Topological insulators and superconductors},\ }\href {https://doi.org/10.1103/RevModPhys.83.1057} {\bibfield  {journal} {\bibinfo  {journal} {Rev. Mod. Phys.}\ }\textbf {\bibinfo {volume} {83}},\ \bibinfo {pages} {1057} (\bibinfo {year} {2011})}\BibitemShut {NoStop}%
\bibitem [{\citenamefont {Bychkov}\ and\ \citenamefont {Rashba}(1984)}]{Rashba_effect2}%
  \BibitemOpen
  \bibfield  {author} {\bibinfo {author} {\bibfnamefont {Y.~A.}\ \bibnamefont {Bychkov}}\ and\ \bibinfo {author} {\bibfnamefont {{\'E}.~I.}\ \bibnamefont {Rashba}},\ }\bibfield  {title} {\bibinfo {title} {{Properties of a 2D electron gas with lifted spectral degeneracy}},\ }\href@noop {} {\bibfield  {journal} {\bibinfo  {journal} {JETP lett}\ }\textbf {\bibinfo {volume} {39}},\ \bibinfo {pages} {78} (\bibinfo {year} {1984})}\BibitemShut {NoStop}%
\bibitem [{\citenamefont {Dresselhaus}(1955)}]{Dresselhaus}%
  \BibitemOpen
  \bibfield  {author} {\bibinfo {author} {\bibfnamefont {G.}~\bibnamefont {Dresselhaus}},\ }\bibfield  {title} {\bibinfo {title} {{Spin-Orbit Coupling Effects in Zinc Blende Structures}},\ }\href {https://doi.org/10.1103/PhysRev.100.580} {\bibfield  {journal} {\bibinfo  {journal} {Phys. Rev.}\ }\textbf {\bibinfo {volume} {100}},\ \bibinfo {pages} {580} (\bibinfo {year} {1955})}\BibitemShut {NoStop}%
\bibitem [{\citenamefont {Mak}\ \emph {et~al.}(2014)\citenamefont {Mak}, \citenamefont {McGill}, \citenamefont {Park},\ and\ \citenamefont {McEuen}}]{2014VHEMak}%
  \BibitemOpen
  \bibfield  {author} {\bibinfo {author} {\bibfnamefont {K.~F.}\ \bibnamefont {Mak}}, \bibinfo {author} {\bibfnamefont {K.~L.}\ \bibnamefont {McGill}}, \bibinfo {author} {\bibfnamefont {J.}~\bibnamefont {Park}},\ and\ \bibinfo {author} {\bibfnamefont {P.~L.}\ \bibnamefont {McEuen}},\ }\bibfield  {title} {\bibinfo {title} {{The valley Hall effect in MoS$_2$ transistors}},\ }\href {https://doi.org/10.1126/science.1250140} {\bibfield  {journal} {\bibinfo  {journal} {Science}\ }\textbf {\bibinfo {volume} {344}},\ \bibinfo {pages} {1489} (\bibinfo {year} {2014})}\BibitemShut {NoStop}%
\bibitem [{\citenamefont {Sinova}\ \emph {et~al.}(2004)\citenamefont {Sinova}, \citenamefont {Culcer}, \citenamefont {Niu}, \citenamefont {Sinitsyn}, \citenamefont {Jungwirth},\ and\ \citenamefont {MacDonald}}]{SHE}%
  \BibitemOpen
  \bibfield  {author} {\bibinfo {author} {\bibfnamefont {J.}~\bibnamefont {Sinova}}, \bibinfo {author} {\bibfnamefont {D.}~\bibnamefont {Culcer}}, \bibinfo {author} {\bibfnamefont {Q.}~\bibnamefont {Niu}}, \bibinfo {author} {\bibfnamefont {N.~A.}\ \bibnamefont {Sinitsyn}}, \bibinfo {author} {\bibfnamefont {T.}~\bibnamefont {Jungwirth}},\ and\ \bibinfo {author} {\bibfnamefont {A.~H.}\ \bibnamefont {MacDonald}},\ }\bibfield  {title} {\bibinfo {title} {{Universal Intrinsic Spin Hall Effect}},\ }\href {https://doi.org/10.1103/PhysRevLett.92.126603} {\bibfield  {journal} {\bibinfo  {journal} {Phys. Rev. Lett.}\ }\textbf {\bibinfo {volume} {92}},\ \bibinfo {pages} {126603} (\bibinfo {year} {2004})}\BibitemShut {NoStop}%
\bibitem [{\citenamefont {Safeer}\ \emph {et~al.}(2019)\citenamefont {Safeer}, \citenamefont {Ingla-Ayn{\'e}s}, \citenamefont {Herling}, \citenamefont {Garcia}, \citenamefont {Vila}, \citenamefont {Ontoso}, \citenamefont {Calvo}, \citenamefont {Roche}, \citenamefont {Hueso},\ and\ \citenamefont {Casanova}}]{SH_mos22019room}%
  \BibitemOpen
  \bibfield  {author} {\bibinfo {author} {\bibfnamefont {C.~K.}\ \bibnamefont {Safeer}}, \bibinfo {author} {\bibfnamefont {J.}~\bibnamefont {Ingla-Ayn{\'e}s}}, \bibinfo {author} {\bibfnamefont {F.}~\bibnamefont {Herling}}, \bibinfo {author} {\bibfnamefont {J.~H.}\ \bibnamefont {Garcia}}, \bibinfo {author} {\bibfnamefont {M.}~\bibnamefont {Vila}}, \bibinfo {author} {\bibfnamefont {N.}~\bibnamefont {Ontoso}}, \bibinfo {author} {\bibfnamefont {M.~R.}\ \bibnamefont {Calvo}}, \bibinfo {author} {\bibfnamefont {S.}~\bibnamefont {Roche}}, \bibinfo {author} {\bibfnamefont {L.~E.}\ \bibnamefont {Hueso}},\ and\ \bibinfo {author} {\bibfnamefont {F.}~\bibnamefont {Casanova}},\ }\bibfield  {title} {\bibinfo {title} {{Room-Temperature Spin Hall Effect in Graphene/MoS$_2$ van der Waals Heterostructures}},\ }\href {https://doi.org/10.1021/acs.nanolett.8b04368} {\bibfield  {journal} {\bibinfo  {journal} {Nano Lett.}\ }\textbf {\bibinfo {volume} {19}},\ \bibinfo {pages} {1074} (\bibinfo {year} {2019})}\BibitemShut {NoStop}%
\bibitem [{\citenamefont {Zhang}\ \emph {et~al.}(2017)\citenamefont {Zhang}, \citenamefont {Sun}, \citenamefont {Yang}, \citenamefont {\ifmmode~\check{Z}\else \v{Z}\fi{}elezn\'y}, \citenamefont {Parkin}, \citenamefont {Felser},\ and\ \citenamefont {Yan}}]{SHE_afm2017strong}%
  \BibitemOpen
  \bibfield  {author} {\bibinfo {author} {\bibfnamefont {Y.}~\bibnamefont {Zhang}}, \bibinfo {author} {\bibfnamefont {Y.}~\bibnamefont {Sun}}, \bibinfo {author} {\bibfnamefont {H.}~\bibnamefont {Yang}}, \bibinfo {author} {\bibfnamefont {J.}~\bibnamefont {\ifmmode~\check{Z}\else \v{Z}\fi{}elezn\'y}}, \bibinfo {author} {\bibfnamefont {S.~P.~P.}\ \bibnamefont {Parkin}}, \bibinfo {author} {\bibfnamefont {C.}~\bibnamefont {Felser}},\ and\ \bibinfo {author} {\bibfnamefont {B.}~\bibnamefont {Yan}},\ }\bibfield  {title} {\bibinfo {title} {{Strong anisotropic anomalous Hall effect and spin Hall effect in the chiral antiferromagnetic compounds ${\mathrm{Mn}}_{3}X$ ($X=\mathrm{Ge}$, Sn, Ga, Ir, Rh, and Pt)}},\ }\href {https://doi.org/10.1103/PhysRevB.95.075128} {\bibfield  {journal} {\bibinfo  {journal} {Phys. Rev. B}\ }\textbf {\bibinfo {volume} {95}},\ \bibinfo {pages} {075128} (\bibinfo {year} {2017})}\BibitemShut {NoStop}%
\bibitem [{\citenamefont {Chen}\ \emph {et~al.}(2021)\citenamefont {Chen}, \citenamefont {Shi}, \citenamefont {Shi}, \citenamefont {Fan}, \citenamefont {Song}, \citenamefont {Zhou}, \citenamefont {Bai}, \citenamefont {Liao}, \citenamefont {Zhou}, \citenamefont {Zhang}, \citenamefont {Li}, \citenamefont {Chen}, \citenamefont {Han}, \citenamefont {Jiang}, \citenamefont {Zhu}, \citenamefont {Wu}, \citenamefont {Wang}, \citenamefont {Xue}, \citenamefont {Yang},\ and\ \citenamefont {Pan}}]{SHE_afm2021observation}%
  \BibitemOpen
  \bibfield  {author} {\bibinfo {author} {\bibfnamefont {X.}~\bibnamefont {Chen}}, \bibinfo {author} {\bibfnamefont {S.}~\bibnamefont {Shi}}, \bibinfo {author} {\bibfnamefont {G.}~\bibnamefont {Shi}}, \bibinfo {author} {\bibfnamefont {X.}~\bibnamefont {Fan}}, \bibinfo {author} {\bibfnamefont {C.}~\bibnamefont {Song}}, \bibinfo {author} {\bibfnamefont {X.}~\bibnamefont {Zhou}}, \bibinfo {author} {\bibfnamefont {H.}~\bibnamefont {Bai}}, \bibinfo {author} {\bibfnamefont {L.}~\bibnamefont {Liao}}, \bibinfo {author} {\bibfnamefont {Y.}~\bibnamefont {Zhou}}, \bibinfo {author} {\bibfnamefont {H.}~\bibnamefont {Zhang}}, \bibinfo {author} {\bibfnamefont {A.}~\bibnamefont {Li}}, \bibinfo {author} {\bibfnamefont {Y.}~\bibnamefont {Chen}}, \bibinfo {author} {\bibfnamefont {X.}~\bibnamefont {Han}}, \bibinfo {author} {\bibfnamefont {S.}~\bibnamefont {Jiang}}, \bibinfo {author} {\bibfnamefont {Z.}~\bibnamefont {Zhu}}, \bibinfo {author} {\bibfnamefont {H.}~\bibnamefont {Wu}}, \bibinfo {author} {\bibfnamefont
  {X.}~\bibnamefont {Wang}}, \bibinfo {author} {\bibfnamefont {D.}~\bibnamefont {Xue}}, \bibinfo {author} {\bibfnamefont {H.}~\bibnamefont {Yang}},\ and\ \bibinfo {author} {\bibfnamefont {F.}~\bibnamefont {Pan}},\ }\bibfield  {title} {\bibinfo {title} {Observation of the antiferromagnetic spin hall effect},\ }\href {https://doi.org/10.1038/s41563-021-00946-z} {\bibfield  {journal} {\bibinfo  {journal} {Nat. Mater.}\ }\textbf {\bibinfo {volume} {20}},\ \bibinfo {pages} {800} (\bibinfo {year} {2021})}\BibitemShut {NoStop}%
\bibitem [{\citenamefont {Liu}\ \emph {et~al.}(2012)\citenamefont {Liu}, \citenamefont {Pai}, \citenamefont {Li}, \citenamefont {Tseng}, \citenamefont {Ralph},\ and\ \citenamefont {Buhrman}}]{Spin-torque_SH}%
  \BibitemOpen
  \bibfield  {author} {\bibinfo {author} {\bibfnamefont {L.}~\bibnamefont {Liu}}, \bibinfo {author} {\bibfnamefont {C.-F.}\ \bibnamefont {Pai}}, \bibinfo {author} {\bibfnamefont {Y.}~\bibnamefont {Li}}, \bibinfo {author} {\bibfnamefont {H.~W.}\ \bibnamefont {Tseng}}, \bibinfo {author} {\bibfnamefont {D.~C.}\ \bibnamefont {Ralph}},\ and\ \bibinfo {author} {\bibfnamefont {R.~A.}\ \bibnamefont {Buhrman}},\ }\bibfield  {title} {\bibinfo {title} {{Spin-Torque Switching with the Giant Spin Hall Effect of Tantalum}},\ }\href {https://doi.org/10.1126/science.1218197} {\bibfield  {journal} {\bibinfo  {journal} {Science}\ }\textbf {\bibinfo {volume} {336}},\ \bibinfo {pages} {555} (\bibinfo {year} {2012})}\BibitemShut {NoStop}%
\bibitem [{\citenamefont {Manchon}\ \emph {et~al.}(2019)\citenamefont {Manchon}, \citenamefont {\ifmmode~\check{Z}\else \v{Z}\fi{}elezn\'y}, \citenamefont {Miron}, \citenamefont {Jungwirth}, \citenamefont {Sinova}, \citenamefont {Thiaville}, \citenamefont {Garello},\ and\ \citenamefont {Gambardella}}]{soctorque_manchon2019current}%
  \BibitemOpen
  \bibfield  {author} {\bibinfo {author} {\bibfnamefont {A.}~\bibnamefont {Manchon}}, \bibinfo {author} {\bibfnamefont {J.}~\bibnamefont {\ifmmode~\check{Z}\else \v{Z}\fi{}elezn\'y}}, \bibinfo {author} {\bibfnamefont {I.~M.}\ \bibnamefont {Miron}}, \bibinfo {author} {\bibfnamefont {T.}~\bibnamefont {Jungwirth}}, \bibinfo {author} {\bibfnamefont {J.}~\bibnamefont {Sinova}}, \bibinfo {author} {\bibfnamefont {A.}~\bibnamefont {Thiaville}}, \bibinfo {author} {\bibfnamefont {K.}~\bibnamefont {Garello}},\ and\ \bibinfo {author} {\bibfnamefont {P.}~\bibnamefont {Gambardella}},\ }\bibfield  {title} {\bibinfo {title} {Current-induced spin-orbit torques in ferromagnetic and antiferromagnetic systems},\ }\href {https://doi.org/10.1103/RevModPhys.91.035004} {\bibfield  {journal} {\bibinfo  {journal} {Rev. Mod. Phys.}\ }\textbf {\bibinfo {volume} {91}},\ \bibinfo {pages} {035004} (\bibinfo {year} {2019})}\BibitemShut {NoStop}%
\bibitem [{\citenamefont {Fukami}\ \emph {et~al.}(2016)\citenamefont {Fukami}, \citenamefont {Zhang}, \citenamefont {DuttaGupta}, \citenamefont {Kurenkov},\ and\ \citenamefont {Ohno}}]{soctorque_fukami2016magnetization}%
  \BibitemOpen
  \bibfield  {author} {\bibinfo {author} {\bibfnamefont {S.}~\bibnamefont {Fukami}}, \bibinfo {author} {\bibfnamefont {C.}~\bibnamefont {Zhang}}, \bibinfo {author} {\bibfnamefont {S.}~\bibnamefont {DuttaGupta}}, \bibinfo {author} {\bibfnamefont {A.}~\bibnamefont {Kurenkov}},\ and\ \bibinfo {author} {\bibfnamefont {H.}~\bibnamefont {Ohno}},\ }\bibfield  {title} {\bibinfo {title} {Magnetization switching by spin--orbit torque in an antiferromagnet--ferromagnet bilayer system},\ }\href {https://doi.org/10.1038/nmat4566} {\bibfield  {journal} {\bibinfo  {journal} {Nat. Mater.}\ }\textbf {\bibinfo {volume} {15}},\ \bibinfo {pages} {535} (\bibinfo {year} {2016})}\BibitemShut {NoStop}%
\bibitem [{\citenamefont {Moore}\ and\ \citenamefont {Read}(1991)}]{MZM_1}%
  \BibitemOpen
  \bibfield  {author} {\bibinfo {author} {\bibfnamefont {G.}~\bibnamefont {Moore}}\ and\ \bibinfo {author} {\bibfnamefont {N.}~\bibnamefont {Read}},\ }\bibfield  {title} {\bibinfo {title} {Nonabelions in the fractional quantum hall effect},\ }\href {https://doi.org/https://doi.org/10.1016/0550-3213(91)90407-O} {\bibfield  {journal} {\bibinfo  {journal} {Nucl. Phys. B}\ }\textbf {\bibinfo {volume} {360}},\ \bibinfo {pages} {362} (\bibinfo {year} {1991})}\BibitemShut {NoStop}%
\bibitem [{\citenamefont {Read}\ and\ \citenamefont {Green}(2000)}]{MZM_2}%
  \BibitemOpen
  \bibfield  {author} {\bibinfo {author} {\bibfnamefont {N.}~\bibnamefont {Read}}\ and\ \bibinfo {author} {\bibfnamefont {D.}~\bibnamefont {Green}},\ }\bibfield  {title} {\bibinfo {title} {Paired states of fermions in two dimensions with breaking of parity and time-reversal symmetries and the fractional quantum hall effect},\ }\href {https://doi.org/10.1103/PhysRevB.61.10267} {\bibfield  {journal} {\bibinfo  {journal} {Phys. Rev. B}\ }\textbf {\bibinfo {volume} {61}},\ \bibinfo {pages} {10267} (\bibinfo {year} {2000})}\BibitemShut {NoStop}%
\bibitem [{\citenamefont {Lutchyn}\ \emph {et~al.}(2018)\citenamefont {Lutchyn}, \citenamefont {Bakkers}, \citenamefont {Kouwenhoven}, \citenamefont {Krogstrup}, \citenamefont {Marcus},\ and\ \citenamefont {Oreg}}]{MZM_review}%
  \BibitemOpen
  \bibfield  {author} {\bibinfo {author} {\bibfnamefont {R.~M.}\ \bibnamefont {Lutchyn}}, \bibinfo {author} {\bibfnamefont {E.~P. A.~M.}\ \bibnamefont {Bakkers}}, \bibinfo {author} {\bibfnamefont {L.~P.}\ \bibnamefont {Kouwenhoven}}, \bibinfo {author} {\bibfnamefont {P.}~\bibnamefont {Krogstrup}}, \bibinfo {author} {\bibfnamefont {C.~M.}\ \bibnamefont {Marcus}},\ and\ \bibinfo {author} {\bibfnamefont {Y.}~\bibnamefont {Oreg}},\ }\bibfield  {title} {\bibinfo {title} {Majorana zero modes in superconductor--semiconductor heterostructures},\ }\href {https://doi.org/10.1038/s41578-018-0003-1} {\bibfield  {journal} {\bibinfo  {journal} {Nature Reviews Materials}\ }\textbf {\bibinfo {volume} {3}},\ \bibinfo {pages} {52} (\bibinfo {year} {2018})}\BibitemShut {NoStop}%
\bibitem [{\citenamefont {Ma}\ \emph {et~al.}(2021)\citenamefont {Ma}, \citenamefont {Hu}, \citenamefont {Li}, \citenamefont {Liu}, \citenamefont {Yao}, \citenamefont {Jia},\ and\ \citenamefont {Liu}}]{CSML}%
  \BibitemOpen
  \bibfield  {author} {\bibinfo {author} {\bibfnamefont {H.-Y.}\ \bibnamefont {Ma}}, \bibinfo {author} {\bibfnamefont {M.}~\bibnamefont {Hu}}, \bibinfo {author} {\bibfnamefont {N.}~\bibnamefont {Li}}, \bibinfo {author} {\bibfnamefont {J.}~\bibnamefont {Liu}}, \bibinfo {author} {\bibfnamefont {W.}~\bibnamefont {Yao}}, \bibinfo {author} {\bibfnamefont {J.-F.}\ \bibnamefont {Jia}},\ and\ \bibinfo {author} {\bibfnamefont {J.}~\bibnamefont {Liu}},\ }\bibfield  {title} {\bibinfo {title} {Multifunctional antiferromagnetic materials with giant piezomagnetism and noncollinear spin current},\ }\href {https://doi.org/10.1038/s41467-021-23127-7} {\bibfield  {journal} {\bibinfo  {journal} {Nat. Commun.}\ }\textbf {\bibinfo {volume} {12}},\ \bibinfo {pages} {2846} (\bibinfo {year} {2021})}\BibitemShut {NoStop}%
\bibitem [{\citenamefont {Wu}\ \emph {et~al.}(2007)\citenamefont {Wu}, \citenamefont {Sun}, \citenamefont {Fradkin},\ and\ \citenamefont {Zhang}}]{spincurrent_zhang}%
  \BibitemOpen
  \bibfield  {author} {\bibinfo {author} {\bibfnamefont {C.}~\bibnamefont {Wu}}, \bibinfo {author} {\bibfnamefont {K.}~\bibnamefont {Sun}}, \bibinfo {author} {\bibfnamefont {E.}~\bibnamefont {Fradkin}},\ and\ \bibinfo {author} {\bibfnamefont {S.-C.}\ \bibnamefont {Zhang}},\ }\bibfield  {title} {\bibinfo {title} {Fermi liquid instabilities in the spin channel},\ }\href {https://doi.org/10.1103/PhysRevB.75.115103} {\bibfield  {journal} {\bibinfo  {journal} {Phys. Rev. B}\ }\textbf {\bibinfo {volume} {75}},\ \bibinfo {pages} {115103} (\bibinfo {year} {2007})}\BibitemShut {NoStop}%
\bibitem [{\citenamefont {Yuan}\ \emph {et~al.}(2020{\natexlab{a}})\citenamefont {Yuan}, \citenamefont {Wang}, \citenamefont {Luo}, \citenamefont {Rashba},\ and\ \citenamefont {Zunger}}]{yuan2020giant}%
  \BibitemOpen
  \bibfield  {author} {\bibinfo {author} {\bibfnamefont {L.-D.}\ \bibnamefont {Yuan}}, \bibinfo {author} {\bibfnamefont {Z.}~\bibnamefont {Wang}}, \bibinfo {author} {\bibfnamefont {J.-W.}\ \bibnamefont {Luo}}, \bibinfo {author} {\bibfnamefont {E.~I.}\ \bibnamefont {Rashba}},\ and\ \bibinfo {author} {\bibfnamefont {A.}~\bibnamefont {Zunger}},\ }\bibfield  {title} {\bibinfo {title} {{Giant momentum-dependent spin splitting in centrosymmetric low-$Z$ antiferromagnets}},\ }\href {https://doi.org/10.1103/PhysRevB.102.014422} {\bibfield  {journal} {\bibinfo  {journal} {Phys. Rev. B}\ }\textbf {\bibinfo {volume} {102}},\ \bibinfo {pages} {014422} (\bibinfo {year} {2020}{\natexlab{a}})}\BibitemShut {NoStop}%
\bibitem [{\citenamefont {Hayami}\ \emph {et~al.}(2019)\citenamefont {Hayami}, \citenamefont {Yanagi},\ and\ \citenamefont {Kusunose}}]{CollinearAFM_splitting}%
  \BibitemOpen
  \bibfield  {author} {\bibinfo {author} {\bibfnamefont {S.}~\bibnamefont {Hayami}}, \bibinfo {author} {\bibfnamefont {Y.}~\bibnamefont {Yanagi}},\ and\ \bibinfo {author} {\bibfnamefont {H.}~\bibnamefont {Kusunose}},\ }\bibfield  {title} {\bibinfo {title} {Momentum-dependent spin splitting by collinear antiferromagnetic ordering},\ }\href {https://doi.org/10.7566/JPSJ.88.123702} {\bibfield  {journal} {\bibinfo  {journal} {J. Phys. Soc. Jpn.}\ }\textbf {\bibinfo {volume} {88}},\ \bibinfo {pages} {123702} (\bibinfo {year} {2019})}\BibitemShut {NoStop}%
\bibitem [{\citenamefont {\ifmmode~\check{S}\else \v{S}\fi{}mejkal}\ \emph {et~al.}(2022{\natexlab{a}})\citenamefont {\ifmmode~\check{S}\else \v{S}\fi{}mejkal}, \citenamefont {Sinova},\ and\ \citenamefont {Jungwirth}}]{sp_sinova2021altermagnetism}%
  \BibitemOpen
  \bibfield  {author} {\bibinfo {author} {\bibfnamefont {L.}~\bibnamefont {\ifmmode~\check{S}\else \v{S}\fi{}mejkal}}, \bibinfo {author} {\bibfnamefont {J.}~\bibnamefont {Sinova}},\ and\ \bibinfo {author} {\bibfnamefont {T.}~\bibnamefont {Jungwirth}},\ }\bibfield  {title} {\bibinfo {title} {{Emerging Research Landscape of Altermagnetism}},\ }\href {https://doi.org/10.1103/PhysRevX.12.040501} {\bibfield  {journal} {\bibinfo  {journal} {Phys. Rev. X}\ }\textbf {\bibinfo {volume} {12}},\ \bibinfo {pages} {040501} (\bibinfo {year} {2022}{\natexlab{a}})}\BibitemShut {NoStop}%
\bibitem [{\citenamefont {\ifmmode~\check{S}\else \v{S}\fi{}mejkal}\ \emph {et~al.}(2022{\natexlab{b}})\citenamefont {\ifmmode~\check{S}\else \v{S}\fi{}mejkal}, \citenamefont {Sinova},\ and\ \citenamefont {Jungwirth}}]{sinova2021altermagnetism2}%
  \BibitemOpen
  \bibfield  {author} {\bibinfo {author} {\bibfnamefont {L.}~\bibnamefont {\ifmmode~\check{S}\else \v{S}\fi{}mejkal}}, \bibinfo {author} {\bibfnamefont {J.}~\bibnamefont {Sinova}},\ and\ \bibinfo {author} {\bibfnamefont {T.}~\bibnamefont {Jungwirth}},\ }\bibfield  {title} {\bibinfo {title} {{Beyond Conventional Ferromagnetism and Antiferromagnetism: A Phase with Nonrelativistic Spin and Crystal Rotation Symmetry}},\ }\href {https://doi.org/10.1103/PhysRevX.12.031042} {\bibfield  {journal} {\bibinfo  {journal} {Phys. Rev. X}\ }\textbf {\bibinfo {volume} {12}},\ \bibinfo {pages} {031042} (\bibinfo {year} {2022}{\natexlab{b}})}\BibitemShut {NoStop}%
\bibitem [{\citenamefont {Bai}\ \emph {et~al.}(2024)\citenamefont {Bai}, \citenamefont {Feng}, \citenamefont {Liu}, \citenamefont {{\v{S}}mejkal}, \citenamefont {Mokrousov},\ and\ \citenamefont {Yao}}]{bai2024altermagnetism}%
  \BibitemOpen
  \bibfield  {author} {\bibinfo {author} {\bibfnamefont {L.}~\bibnamefont {Bai}}, \bibinfo {author} {\bibfnamefont {W.}~\bibnamefont {Feng}}, \bibinfo {author} {\bibfnamefont {S.}~\bibnamefont {Liu}}, \bibinfo {author} {\bibfnamefont {L.}~\bibnamefont {{\v{S}}mejkal}}, \bibinfo {author} {\bibfnamefont {Y.}~\bibnamefont {Mokrousov}},\ and\ \bibinfo {author} {\bibfnamefont {Y.}~\bibnamefont {Yao}},\ }\bibfield  {title} {\bibinfo {title} {Altermagnetism: Exploring new frontiers in magnetism and spintronics},\ }\href@noop {} {\bibfield  {journal} {\bibinfo  {journal} {Advanced Functional Materials}\ }\textbf {\bibinfo {volume} {34}},\ \bibinfo {pages} {2409327} (\bibinfo {year} {2024})}\BibitemShut {NoStop}%
\bibitem [{\citenamefont {Song}\ \emph {et~al.}(2025)\citenamefont {Song}, \citenamefont {Bai}, \citenamefont {Zhou}, \citenamefont {Han}, \citenamefont {Reichlova}, \citenamefont {Dil}, \citenamefont {Liu}, \citenamefont {Chen},\ and\ \citenamefont {Pan}}]{Song2025review}%
  \BibitemOpen
  \bibfield  {author} {\bibinfo {author} {\bibfnamefont {C.}~\bibnamefont {Song}}, \bibinfo {author} {\bibfnamefont {H.}~\bibnamefont {Bai}}, \bibinfo {author} {\bibfnamefont {Z.}~\bibnamefont {Zhou}}, \bibinfo {author} {\bibfnamefont {L.}~\bibnamefont {Han}}, \bibinfo {author} {\bibfnamefont {H.}~\bibnamefont {Reichlova}}, \bibinfo {author} {\bibfnamefont {J.~H.}\ \bibnamefont {Dil}}, \bibinfo {author} {\bibfnamefont {J.}~\bibnamefont {Liu}}, \bibinfo {author} {\bibfnamefont {X.}~\bibnamefont {Chen}},\ and\ \bibinfo {author} {\bibfnamefont {F.}~\bibnamefont {Pan}},\ }\bibfield  {title} {\bibinfo {title} {Altermagnets as a new class of functional materials},\ }\bibfield  {journal} {\bibinfo  {journal} {Nature Reviews Materials}\ }\href {https://doi.org/10.1038/s41578-025-00779-1} {10.1038/s41578-025-00779-1} (\bibinfo {year} {2025})\BibitemShut {NoStop}%
\bibitem [{\citenamefont {Gonz\'alez-Hern\'andez}\ \emph {et~al.}(2021)\citenamefont {Gonz\'alez-Hern\'andez}, \citenamefont {\ifmmode~\check{S}\else \v{S}\fi{}mejkal}, \citenamefont {V\'yborn\'y}, \citenamefont {Yahagi}, \citenamefont {Sinova}, \citenamefont {Jungwirth},\ and\ \citenamefont {\ifmmode~\check{Z}\else \v{Z}\fi{}elezn\'y}}]{spinsplitter_libor}%
  \BibitemOpen
  \bibfield  {author} {\bibinfo {author} {\bibfnamefont {R.}~\bibnamefont {Gonz\'alez-Hern\'andez}}, \bibinfo {author} {\bibfnamefont {L.}~\bibnamefont {\ifmmode~\check{S}\else \v{S}\fi{}mejkal}}, \bibinfo {author} {\bibfnamefont {K.}~\bibnamefont {V\'yborn\'y}}, \bibinfo {author} {\bibfnamefont {Y.}~\bibnamefont {Yahagi}}, \bibinfo {author} {\bibfnamefont {J.}~\bibnamefont {Sinova}}, \bibinfo {author} {\bibfnamefont {T.~c.~v.}\ \bibnamefont {Jungwirth}},\ and\ \bibinfo {author} {\bibfnamefont {J.}~\bibnamefont {\ifmmode~\check{Z}\else \v{Z}\fi{}elezn\'y}},\ }\bibfield  {title} {\bibinfo {title} {{Efficient Electrical Spin Splitter Based on Nonrelativistic Collinear Antiferromagnetism}},\ }\href {https://doi.org/10.1103/PhysRevLett.126.127701} {\bibfield  {journal} {\bibinfo  {journal} {Phys. Rev. Lett.}\ }\textbf {\bibinfo {volume} {126}},\ \bibinfo {pages} {127701} (\bibinfo {year} {2021})}\BibitemShut {NoStop}%
\bibitem [{\citenamefont {Naka}\ \emph {et~al.}(2019)\citenamefont {Naka}, \citenamefont {Hayami}, \citenamefont {Kusunose}, \citenamefont {Yanagi}, \citenamefont {Motome},\ and\ \citenamefont {Seo}}]{Naka2019}%
  \BibitemOpen
  \bibfield  {author} {\bibinfo {author} {\bibfnamefont {M.}~\bibnamefont {Naka}}, \bibinfo {author} {\bibfnamefont {S.}~\bibnamefont {Hayami}}, \bibinfo {author} {\bibfnamefont {H.}~\bibnamefont {Kusunose}}, \bibinfo {author} {\bibfnamefont {Y.}~\bibnamefont {Yanagi}}, \bibinfo {author} {\bibfnamefont {Y.}~\bibnamefont {Motome}},\ and\ \bibinfo {author} {\bibfnamefont {H.}~\bibnamefont {Seo}},\ }\bibfield  {title} {\bibinfo {title} {Spin current generation in organic antiferromagnets},\ }\href {https://doi.org/10.1038/s41467-019-12229-y} {\bibfield  {journal} {\bibinfo  {journal} {Nat. Commun.}\ }\textbf {\bibinfo {volume} {10}},\ \bibinfo {pages} {4305} (\bibinfo {year} {2019})}\BibitemShut {NoStop}%
\bibitem [{\citenamefont {Aoyama}\ and\ \citenamefont {Ohgushi}(2024)}]{PZM_MnTe}%
  \BibitemOpen
  \bibfield  {author} {\bibinfo {author} {\bibfnamefont {T.}~\bibnamefont {Aoyama}}\ and\ \bibinfo {author} {\bibfnamefont {K.}~\bibnamefont {Ohgushi}},\ }\bibfield  {title} {\bibinfo {title} {{Piezomagnetic properties in altermagnetic MnTe}},\ }\href {https://doi.org/10.1103/PhysRevMaterials.8.L041402} {\bibfield  {journal} {\bibinfo  {journal} {Phys. Rev. Mater.}\ }\textbf {\bibinfo {volume} {8}},\ \bibinfo {pages} {L041402} (\bibinfo {year} {2024})}\BibitemShut {NoStop}%
\bibitem [{\citenamefont {Bose}\ \emph {et~al.}(2022)\citenamefont {Bose}, \citenamefont {Schreiber}, \citenamefont {Jain}, \citenamefont {Shao}, \citenamefont {Nair}, \citenamefont {Sun}, \citenamefont {Zhang}, \citenamefont {Muller}, \citenamefont {Tsymbal}, \citenamefont {Schlom},\ and\ \citenamefont {Ralph}}]{ruo2_spincurrent}%
  \BibitemOpen
  \bibfield  {author} {\bibinfo {author} {\bibfnamefont {A.}~\bibnamefont {Bose}}, \bibinfo {author} {\bibfnamefont {N.~J.}\ \bibnamefont {Schreiber}}, \bibinfo {author} {\bibfnamefont {R.}~\bibnamefont {Jain}}, \bibinfo {author} {\bibfnamefont {D.-F.}\ \bibnamefont {Shao}}, \bibinfo {author} {\bibfnamefont {H.~P.}\ \bibnamefont {Nair}}, \bibinfo {author} {\bibfnamefont {J.}~\bibnamefont {Sun}}, \bibinfo {author} {\bibfnamefont {X.~S.}\ \bibnamefont {Zhang}}, \bibinfo {author} {\bibfnamefont {D.~A.}\ \bibnamefont {Muller}}, \bibinfo {author} {\bibfnamefont {E.~Y.}\ \bibnamefont {Tsymbal}}, \bibinfo {author} {\bibfnamefont {D.~G.}\ \bibnamefont {Schlom}},\ and\ \bibinfo {author} {\bibfnamefont {D.~C.}\ \bibnamefont {Ralph}},\ }\bibfield  {title} {\bibinfo {title} {{Tilted spin current generated by the collinear antiferromagnet ruthenium dioxide}},\ }\href {https://doi.org/10.1038/s41928-022-00744-8} {\bibfield  {journal} {\bibinfo  {journal} {Nat. Electron.}\ }\textbf {\bibinfo {volume} {5}},\ \bibinfo {pages}
  {267} (\bibinfo {year} {2022})}\BibitemShut {NoStop}%
\bibitem [{\citenamefont {Bai}\ \emph {et~al.}(2022)\citenamefont {Bai}, \citenamefont {Han}, \citenamefont {Feng}, \citenamefont {Zhou}, \citenamefont {Su}, \citenamefont {Wang}, \citenamefont {Liao}, \citenamefont {Zhu}, \citenamefont {Chen}, \citenamefont {Pan}, \citenamefont {Fan},\ and\ \citenamefont {Song}}]{torque_bai2021observation}%
  \BibitemOpen
  \bibfield  {author} {\bibinfo {author} {\bibfnamefont {H.}~\bibnamefont {Bai}}, \bibinfo {author} {\bibfnamefont {L.}~\bibnamefont {Han}}, \bibinfo {author} {\bibfnamefont {X.~Y.}\ \bibnamefont {Feng}}, \bibinfo {author} {\bibfnamefont {Y.~J.}\ \bibnamefont {Zhou}}, \bibinfo {author} {\bibfnamefont {R.~X.}\ \bibnamefont {Su}}, \bibinfo {author} {\bibfnamefont {Q.}~\bibnamefont {Wang}}, \bibinfo {author} {\bibfnamefont {L.~Y.}\ \bibnamefont {Liao}}, \bibinfo {author} {\bibfnamefont {W.~X.}\ \bibnamefont {Zhu}}, \bibinfo {author} {\bibfnamefont {X.~Z.}\ \bibnamefont {Chen}}, \bibinfo {author} {\bibfnamefont {F.}~\bibnamefont {Pan}}, \bibinfo {author} {\bibfnamefont {X.~L.}\ \bibnamefont {Fan}},\ and\ \bibinfo {author} {\bibfnamefont {C.}~\bibnamefont {Song}},\ }\bibfield  {title} {\bibinfo {title} {{Observation of Spin Splitting Torque in a Collinear Antiferromagnet RuO$_{2}$}},\ }\href {https://doi.org/10.1103/PhysRevLett.128.197202} {\bibfield  {journal} {\bibinfo  {journal} {Phys. Rev. Lett.}\ }\textbf
  {\bibinfo {volume} {128}},\ \bibinfo {pages} {197202} (\bibinfo {year} {2022})}\BibitemShut {NoStop}%
\bibitem [{\citenamefont {Karube}\ \emph {et~al.}(2022)\citenamefont {Karube}, \citenamefont {Tanaka}, \citenamefont {Sugawara}, \citenamefont {Kadoguchi}, \citenamefont {Kohda},\ and\ \citenamefont {Nitta}}]{torque_karube2}%
  \BibitemOpen
  \bibfield  {author} {\bibinfo {author} {\bibfnamefont {S.}~\bibnamefont {Karube}}, \bibinfo {author} {\bibfnamefont {T.}~\bibnamefont {Tanaka}}, \bibinfo {author} {\bibfnamefont {D.}~\bibnamefont {Sugawara}}, \bibinfo {author} {\bibfnamefont {N.}~\bibnamefont {Kadoguchi}}, \bibinfo {author} {\bibfnamefont {M.}~\bibnamefont {Kohda}},\ and\ \bibinfo {author} {\bibfnamefont {J.}~\bibnamefont {Nitta}},\ }\bibfield  {title} {\bibinfo {title} {{Observation of Spin-Splitter Torque in Collinear Antiferromagnetic ${\mathrm{RuO}}_{2}$}},\ }\href {https://doi.org/10.1103/PhysRevLett.129.137201} {\bibfield  {journal} {\bibinfo  {journal} {Phys. Rev. Lett.}\ }\textbf {\bibinfo {volume} {129}},\ \bibinfo {pages} {137201} (\bibinfo {year} {2022})}\BibitemShut {NoStop}%
\bibitem [{\citenamefont {Berlijn}\ \emph {et~al.}(2017)\citenamefont {Berlijn}, \citenamefont {Snijders}, \citenamefont {Delaire}, \citenamefont {Zhou}, \citenamefont {Maier}, \citenamefont {Cao}, \citenamefont {Chi}, \citenamefont {Matsuda}, \citenamefont {Wang}, \citenamefont {Koehler}, \citenamefont {Kent},\ and\ \citenamefont {Weitering}}]{ruo2_exp}%
  \BibitemOpen
  \bibfield  {author} {\bibinfo {author} {\bibfnamefont {T.}~\bibnamefont {Berlijn}}, \bibinfo {author} {\bibfnamefont {P.~C.}\ \bibnamefont {Snijders}}, \bibinfo {author} {\bibfnamefont {O.}~\bibnamefont {Delaire}}, \bibinfo {author} {\bibfnamefont {H.-D.}\ \bibnamefont {Zhou}}, \bibinfo {author} {\bibfnamefont {T.~A.}\ \bibnamefont {Maier}}, \bibinfo {author} {\bibfnamefont {H.-B.}\ \bibnamefont {Cao}}, \bibinfo {author} {\bibfnamefont {S.-X.}\ \bibnamefont {Chi}}, \bibinfo {author} {\bibfnamefont {M.}~\bibnamefont {Matsuda}}, \bibinfo {author} {\bibfnamefont {Y.}~\bibnamefont {Wang}}, \bibinfo {author} {\bibfnamefont {M.~R.}\ \bibnamefont {Koehler}}, \bibinfo {author} {\bibfnamefont {P.~R.~C.}\ \bibnamefont {Kent}},\ and\ \bibinfo {author} {\bibfnamefont {H.~H.}\ \bibnamefont {Weitering}},\ }\bibfield  {title} {\bibinfo {title} {{Itinerant Antiferromagnetism in ${\mathrm{RuO}}_{2}$}},\ }\href {https://doi.org/10.1103/PhysRevLett.118.077201} {\bibfield  {journal} {\bibinfo  {journal} {Phys. Rev. Lett.}\
  }\textbf {\bibinfo {volume} {118}},\ \bibinfo {pages} {077201} (\bibinfo {year} {2017})}\BibitemShut {NoStop}%
\bibitem [{\citenamefont {Fedchenko}\ \emph {et~al.}(2024)\citenamefont {Fedchenko}, \citenamefont {Minár}, \citenamefont {Akashdeep}, \citenamefont {D’Souza}, \citenamefont {Vasilyev}, \citenamefont {Tkach}, \citenamefont {Odenbreit}, \citenamefont {Nguyen}, \citenamefont {Kutnyakhov}, \citenamefont {Wind}, \citenamefont {Wenthaus}, \citenamefont {Scholz}, \citenamefont {Rossnagel}, \citenamefont {Hoesch}, \citenamefont {Aeschlimann}, \citenamefont {Stadtmüller}, \citenamefont {Kläui}, \citenamefont {Schönhense}, \citenamefont {Jungwirth}, \citenamefont {Hellenes}, \citenamefont {Jakob}, \citenamefont {Šmejkal}, \citenamefont {Sinova},\ and\ \citenamefont {Elmers}}]{RuO2_ARPES}%
  \BibitemOpen
  \bibfield  {author} {\bibinfo {author} {\bibfnamefont {O.}~\bibnamefont {Fedchenko}}, \bibinfo {author} {\bibfnamefont {J.}~\bibnamefont {Minár}}, \bibinfo {author} {\bibfnamefont {A.}~\bibnamefont {Akashdeep}}, \bibinfo {author} {\bibfnamefont {S.~W.}\ \bibnamefont {D’Souza}}, \bibinfo {author} {\bibfnamefont {D.}~\bibnamefont {Vasilyev}}, \bibinfo {author} {\bibfnamefont {O.}~\bibnamefont {Tkach}}, \bibinfo {author} {\bibfnamefont {L.}~\bibnamefont {Odenbreit}}, \bibinfo {author} {\bibfnamefont {Q.}~\bibnamefont {Nguyen}}, \bibinfo {author} {\bibfnamefont {D.}~\bibnamefont {Kutnyakhov}}, \bibinfo {author} {\bibfnamefont {N.}~\bibnamefont {Wind}}, \bibinfo {author} {\bibfnamefont {L.}~\bibnamefont {Wenthaus}}, \bibinfo {author} {\bibfnamefont {M.}~\bibnamefont {Scholz}}, \bibinfo {author} {\bibfnamefont {K.}~\bibnamefont {Rossnagel}}, \bibinfo {author} {\bibfnamefont {M.}~\bibnamefont {Hoesch}}, \bibinfo {author} {\bibfnamefont {M.}~\bibnamefont {Aeschlimann}}, \bibinfo {author} {\bibfnamefont
  {B.}~\bibnamefont {Stadtmüller}}, \bibinfo {author} {\bibfnamefont {M.}~\bibnamefont {Kläui}}, \bibinfo {author} {\bibfnamefont {G.}~\bibnamefont {Schönhense}}, \bibinfo {author} {\bibfnamefont {T.}~\bibnamefont {Jungwirth}}, \bibinfo {author} {\bibfnamefont {A.~B.}\ \bibnamefont {Hellenes}}, \bibinfo {author} {\bibfnamefont {G.}~\bibnamefont {Jakob}}, \bibinfo {author} {\bibfnamefont {L.}~\bibnamefont {Šmejkal}}, \bibinfo {author} {\bibfnamefont {J.}~\bibnamefont {Sinova}},\ and\ \bibinfo {author} {\bibfnamefont {H.-J.}\ \bibnamefont {Elmers}},\ }\bibfield  {title} {\bibinfo {title} {{Observation of time-reversal symmetry breaking in the band structure of altermagnetic RuO$_2$}},\ }\href {https://doi.org/10.1126/sciadv.adj4883} {\bibfield  {journal} {\bibinfo  {journal} {Sci. Adv.}\ }\textbf {\bibinfo {volume} {10}},\ \bibinfo {pages} {eadj4883} (\bibinfo {year} {2024})}\BibitemShut {NoStop}%
\bibitem [{\citenamefont {{Kunitomi, Nobuhiko}}\ \emph {et~al.}(1964)\citenamefont {{Kunitomi, Nobuhiko}}, \citenamefont {{Hamaguchi, Yoshikazu}},\ and\ \citenamefont {{Anzai, Shuichiro}}}]{MnTe_neutron}%
  \BibitemOpen
  \bibfield  {author} {\bibinfo {author} {\bibnamefont {{Kunitomi, Nobuhiko}}}, \bibinfo {author} {\bibnamefont {{Hamaguchi, Yoshikazu}}},\ and\ \bibinfo {author} {\bibnamefont {{Anzai, Shuichiro}}},\ }\bibfield  {title} {\bibinfo {title} {{Neutron diffraction study on manganese telluride}},\ }\href {https://doi.org/10.1051/jphys:01964002505056800} {\bibfield  {journal} {\bibinfo  {journal} {J. Phys. France}\ }\textbf {\bibinfo {volume} {25}},\ \bibinfo {pages} {568} (\bibinfo {year} {1964})}\BibitemShut {NoStop}%
\bibitem [{\citenamefont {Krempask{\'y}}\ \emph {et~al.}(2024)\citenamefont {Krempask{\'y}}, \citenamefont {{\v{S}}mejkal}, \citenamefont {D'Souza}, \citenamefont {Hajlaoui}, \citenamefont {Springholz}, \citenamefont {Uhl{\'i}{\v{r}}ov{\'a}}, \citenamefont {Alarab}, \citenamefont {Constantinou}, \citenamefont {Strocov}, \citenamefont {Usanov}, \citenamefont {Pudelko}, \citenamefont {Gonz{\'a}lez-Hern{\'a}ndez}, \citenamefont {Birk~Hellenes}, \citenamefont {Jansa}, \citenamefont {Reichlov{\'a}}, \citenamefont {{\v{S}}ob{\'a}{\v{n}}}, \citenamefont {Gonzalez~Betancourt}, \citenamefont {Wadley}, \citenamefont {Sinova}, \citenamefont {Kriegner}, \citenamefont {Min{\'a}r}, \citenamefont {Dil},\ and\ \citenamefont {Jungwirth}}]{MnTe_ARPES}%
  \BibitemOpen
  \bibfield  {author} {\bibinfo {author} {\bibfnamefont {J.}~\bibnamefont {Krempask{\'y}}}, \bibinfo {author} {\bibfnamefont {L.}~\bibnamefont {{\v{S}}mejkal}}, \bibinfo {author} {\bibfnamefont {S.~W.}\ \bibnamefont {D'Souza}}, \bibinfo {author} {\bibfnamefont {M.}~\bibnamefont {Hajlaoui}}, \bibinfo {author} {\bibfnamefont {G.}~\bibnamefont {Springholz}}, \bibinfo {author} {\bibfnamefont {K.}~\bibnamefont {Uhl{\'i}{\v{r}}ov{\'a}}}, \bibinfo {author} {\bibfnamefont {F.}~\bibnamefont {Alarab}}, \bibinfo {author} {\bibfnamefont {P.~C.}\ \bibnamefont {Constantinou}}, \bibinfo {author} {\bibfnamefont {V.}~\bibnamefont {Strocov}}, \bibinfo {author} {\bibfnamefont {D.}~\bibnamefont {Usanov}}, \bibinfo {author} {\bibfnamefont {W.~R.}\ \bibnamefont {Pudelko}}, \bibinfo {author} {\bibfnamefont {R.}~\bibnamefont {Gonz{\'a}lez-Hern{\'a}ndez}}, \bibinfo {author} {\bibfnamefont {A.}~\bibnamefont {Birk~Hellenes}}, \bibinfo {author} {\bibfnamefont {Z.}~\bibnamefont {Jansa}}, \bibinfo {author} {\bibfnamefont {H.}~\bibnamefont
  {Reichlov{\'a}}}, \bibinfo {author} {\bibfnamefont {Z.}~\bibnamefont {{\v{S}}ob{\'a}{\v{n}}}}, \bibinfo {author} {\bibfnamefont {R.~D.}\ \bibnamefont {Gonzalez~Betancourt}}, \bibinfo {author} {\bibfnamefont {P.}~\bibnamefont {Wadley}}, \bibinfo {author} {\bibfnamefont {J.}~\bibnamefont {Sinova}}, \bibinfo {author} {\bibfnamefont {D.}~\bibnamefont {Kriegner}}, \bibinfo {author} {\bibfnamefont {J.}~\bibnamefont {Min{\'a}r}}, \bibinfo {author} {\bibfnamefont {J.~H.}\ \bibnamefont {Dil}},\ and\ \bibinfo {author} {\bibfnamefont {T.}~\bibnamefont {Jungwirth}},\ }\bibfield  {title} {\bibinfo {title} {Altermagnetic lifting of kramers spin degeneracy},\ }\href {https://doi.org/10.1038/s41586-023-06907-7} {\bibfield  {journal} {\bibinfo  {journal} {Nature}\ }\textbf {\bibinfo {volume} {626}},\ \bibinfo {pages} {517} (\bibinfo {year} {2024})}\BibitemShut {NoStop}%
\bibitem [{\citenamefont {Yuan}\ \emph {et~al.}(2020{\natexlab{b}})\citenamefont {Yuan}, \citenamefont {Song}, \citenamefont {Xing},\ and\ \citenamefont {Chen}}]{CrSb_crystal}%
  \BibitemOpen
  \bibfield  {author} {\bibinfo {author} {\bibfnamefont {J.}~\bibnamefont {Yuan}}, \bibinfo {author} {\bibfnamefont {Y.}~\bibnamefont {Song}}, \bibinfo {author} {\bibfnamefont {X.}~\bibnamefont {Xing}},\ and\ \bibinfo {author} {\bibfnamefont {J.}~\bibnamefont {Chen}},\ }\bibfield  {title} {\bibinfo {title} {{Magnetic structure and uniaxial negative thermal expansion in antiferromagnetic CrSb}},\ }\href {https://doi.org/10.1039/D0DT03277H} {\bibfield  {journal} {\bibinfo  {journal} {Dalton Trans.}\ }\textbf {\bibinfo {volume} {49}},\ \bibinfo {pages} {17605} (\bibinfo {year} {2020}{\natexlab{b}})}\BibitemShut {NoStop}%
\bibitem [{\citenamefont {Reimers}\ \emph {et~al.}(2024)\citenamefont {Reimers}, \citenamefont {Odenbreit}, \citenamefont {{\v{S}}mejkal}, \citenamefont {Strocov}, \citenamefont {Constantinou}, \citenamefont {Hellenes}, \citenamefont {Jaeschke~Ubiergo}, \citenamefont {Campos}, \citenamefont {Bharadwaj}, \citenamefont {Chakraborty}, \citenamefont {Denneulin}, \citenamefont {Shi}, \citenamefont {Dunin-Borkowski}, \citenamefont {Das}, \citenamefont {Kl{\"a}ui}, \citenamefont {Sinova},\ and\ \citenamefont {Jourdan}}]{CrSb_APRES}%
  \BibitemOpen
  \bibfield  {author} {\bibinfo {author} {\bibfnamefont {S.}~\bibnamefont {Reimers}}, \bibinfo {author} {\bibfnamefont {L.}~\bibnamefont {Odenbreit}}, \bibinfo {author} {\bibfnamefont {L.}~\bibnamefont {{\v{S}}mejkal}}, \bibinfo {author} {\bibfnamefont {V.~N.}\ \bibnamefont {Strocov}}, \bibinfo {author} {\bibfnamefont {P.}~\bibnamefont {Constantinou}}, \bibinfo {author} {\bibfnamefont {A.~B.}\ \bibnamefont {Hellenes}}, \bibinfo {author} {\bibfnamefont {R.}~\bibnamefont {Jaeschke~Ubiergo}}, \bibinfo {author} {\bibfnamefont {W.~H.}\ \bibnamefont {Campos}}, \bibinfo {author} {\bibfnamefont {V.~K.}\ \bibnamefont {Bharadwaj}}, \bibinfo {author} {\bibfnamefont {A.}~\bibnamefont {Chakraborty}}, \bibinfo {author} {\bibfnamefont {T.}~\bibnamefont {Denneulin}}, \bibinfo {author} {\bibfnamefont {W.}~\bibnamefont {Shi}}, \bibinfo {author} {\bibfnamefont {R.~E.}\ \bibnamefont {Dunin-Borkowski}}, \bibinfo {author} {\bibfnamefont {S.}~\bibnamefont {Das}}, \bibinfo {author} {\bibfnamefont {M.}~\bibnamefont {Kl{\"a}ui}},
  \bibinfo {author} {\bibfnamefont {J.}~\bibnamefont {Sinova}},\ and\ \bibinfo {author} {\bibfnamefont {M.}~\bibnamefont {Jourdan}},\ }\bibfield  {title} {\bibinfo {title} {Direct observation of altermagnetic band splitting in crsb thin films},\ }\href {https://doi.org/10.1038/s41467-024-46476-5} {\bibfield  {journal} {\bibinfo  {journal} {Nat. Commun.}\ }\textbf {\bibinfo {volume} {15}},\ \bibinfo {pages} {2116} (\bibinfo {year} {2024})}\BibitemShut {NoStop}%
\bibitem [{\citenamefont {Li}\ \emph {et~al.}(2024)\citenamefont {Li}, \citenamefont {Hu}, \citenamefont {Li}, \citenamefont {Wang}, \citenamefont {Chen}, \citenamefont {Thiagarajan}, \citenamefont {Leandersson}, \citenamefont {Polley}, \citenamefont {Kim}, \citenamefont {Liu}, \citenamefont {Fulga}, \citenamefont {Vergniory}, \citenamefont {Janson}, \citenamefont {Tjernberg},\ and\ \citenamefont {van~den Brink}}]{CrSb_topological}%
  \BibitemOpen
  \bibfield  {author} {\bibinfo {author} {\bibfnamefont {C.}~\bibnamefont {Li}}, \bibinfo {author} {\bibfnamefont {M.}~\bibnamefont {Hu}}, \bibinfo {author} {\bibfnamefont {Z.}~\bibnamefont {Li}}, \bibinfo {author} {\bibfnamefont {Y.}~\bibnamefont {Wang}}, \bibinfo {author} {\bibfnamefont {W.}~\bibnamefont {Chen}}, \bibinfo {author} {\bibfnamefont {B.}~\bibnamefont {Thiagarajan}}, \bibinfo {author} {\bibfnamefont {M.}~\bibnamefont {Leandersson}}, \bibinfo {author} {\bibfnamefont {C.}~\bibnamefont {Polley}}, \bibinfo {author} {\bibfnamefont {T.}~\bibnamefont {Kim}}, \bibinfo {author} {\bibfnamefont {H.}~\bibnamefont {Liu}}, \bibinfo {author} {\bibfnamefont {C.}~\bibnamefont {Fulga}}, \bibinfo {author} {\bibfnamefont {M.~G.}\ \bibnamefont {Vergniory}}, \bibinfo {author} {\bibfnamefont {O.}~\bibnamefont {Janson}}, \bibinfo {author} {\bibfnamefont {O.}~\bibnamefont {Tjernberg}},\ and\ \bibinfo {author} {\bibfnamefont {J.}~\bibnamefont {van~den Brink}},\ }\href@noop {} {\bibinfo {title} {{Topological Weyl
  Altermagnetism in CrSb}}} (\bibinfo {year} {2024}),\ \Eprint {https://arxiv.org/abs/2405.14777} {arXiv:2405.14777 [cond-mat.mtrl-sci]} \BibitemShut {NoStop}%
\bibitem [{\citenamefont {Ding}\ \emph {et~al.}(2024)\citenamefont {Ding}, \citenamefont {Jiang}, \citenamefont {Chen}, \citenamefont {Tao}, \citenamefont {Liu}, \citenamefont {Liu}, \citenamefont {Li}, \citenamefont {Liu}, \citenamefont {Yang}, \citenamefont {Zhang}, \citenamefont {Deng}, \citenamefont {Jing}, \citenamefont {Huang}, \citenamefont {Shi}, \citenamefont {Qiao}, \citenamefont {Wang}, \citenamefont {Guo}, \citenamefont {Feng},\ and\ \citenamefont {Shen}}]{CrSb_ARPES2}%
  \BibitemOpen
  \bibfield  {author} {\bibinfo {author} {\bibfnamefont {J.}~\bibnamefont {Ding}}, \bibinfo {author} {\bibfnamefont {Z.}~\bibnamefont {Jiang}}, \bibinfo {author} {\bibfnamefont {X.}~\bibnamefont {Chen}}, \bibinfo {author} {\bibfnamefont {Z.}~\bibnamefont {Tao}}, \bibinfo {author} {\bibfnamefont {Z.}~\bibnamefont {Liu}}, \bibinfo {author} {\bibfnamefont {J.}~\bibnamefont {Liu}}, \bibinfo {author} {\bibfnamefont {T.}~\bibnamefont {Li}}, \bibinfo {author} {\bibfnamefont {J.}~\bibnamefont {Liu}}, \bibinfo {author} {\bibfnamefont {Y.}~\bibnamefont {Yang}}, \bibinfo {author} {\bibfnamefont {R.}~\bibnamefont {Zhang}}, \bibinfo {author} {\bibfnamefont {L.}~\bibnamefont {Deng}}, \bibinfo {author} {\bibfnamefont {W.}~\bibnamefont {Jing}}, \bibinfo {author} {\bibfnamefont {Y.}~\bibnamefont {Huang}}, \bibinfo {author} {\bibfnamefont {Y.}~\bibnamefont {Shi}}, \bibinfo {author} {\bibfnamefont {S.}~\bibnamefont {Qiao}}, \bibinfo {author} {\bibfnamefont {Y.}~\bibnamefont {Wang}}, \bibinfo {author} {\bibfnamefont
  {Y.}~\bibnamefont {Guo}}, \bibinfo {author} {\bibfnamefont {D.}~\bibnamefont {Feng}},\ and\ \bibinfo {author} {\bibfnamefont {D.}~\bibnamefont {Shen}},\ }\href@noop {} {\bibinfo {title} {{Large band-splitting in $g$-wave type altermagnet CrSb}}} (\bibinfo {year} {2024}),\ \Eprint {https://arxiv.org/abs/2405.12687} {arXiv:2405.12687 [cond-mat.mtrl-sci]} \BibitemShut {NoStop}%
\bibitem [{\citenamefont {Zeng}\ \emph {et~al.}(2024)\citenamefont {Zeng}, \citenamefont {Zhu}, \citenamefont {Zhu}, \citenamefont {Liu}, \citenamefont {Ma}, \citenamefont {Hao}, \citenamefont {Liu}, \citenamefont {Qu}, \citenamefont {Yang}, \citenamefont {Jiang}, \citenamefont {Yamagami}, \citenamefont {Arita}, \citenamefont {Zhang}, \citenamefont {Shao}, \citenamefont {Dai}, \citenamefont {Shimada}, \citenamefont {Liu}, \citenamefont {Ye}, \citenamefont {Huang}, \citenamefont {Liu},\ and\ \citenamefont {Liu}}]{CrSb_ARPES3}%
  \BibitemOpen
  \bibfield  {author} {\bibinfo {author} {\bibfnamefont {M.}~\bibnamefont {Zeng}}, \bibinfo {author} {\bibfnamefont {M.-Y.}\ \bibnamefont {Zhu}}, \bibinfo {author} {\bibfnamefont {Y.-P.}\ \bibnamefont {Zhu}}, \bibinfo {author} {\bibfnamefont {X.-R.}\ \bibnamefont {Liu}}, \bibinfo {author} {\bibfnamefont {X.-M.}\ \bibnamefont {Ma}}, \bibinfo {author} {\bibfnamefont {Y.-J.}\ \bibnamefont {Hao}}, \bibinfo {author} {\bibfnamefont {P.}~\bibnamefont {Liu}}, \bibinfo {author} {\bibfnamefont {G.}~\bibnamefont {Qu}}, \bibinfo {author} {\bibfnamefont {Y.}~\bibnamefont {Yang}}, \bibinfo {author} {\bibfnamefont {Z.}~\bibnamefont {Jiang}}, \bibinfo {author} {\bibfnamefont {K.}~\bibnamefont {Yamagami}}, \bibinfo {author} {\bibfnamefont {M.}~\bibnamefont {Arita}}, \bibinfo {author} {\bibfnamefont {X.}~\bibnamefont {Zhang}}, \bibinfo {author} {\bibfnamefont {T.-H.}\ \bibnamefont {Shao}}, \bibinfo {author} {\bibfnamefont {Y.}~\bibnamefont {Dai}}, \bibinfo {author} {\bibfnamefont {K.}~\bibnamefont {Shimada}}, \bibinfo {author}
  {\bibfnamefont {Z.}~\bibnamefont {Liu}}, \bibinfo {author} {\bibfnamefont {M.}~\bibnamefont {Ye}}, \bibinfo {author} {\bibfnamefont {Y.}~\bibnamefont {Huang}}, \bibinfo {author} {\bibfnamefont {Q.}~\bibnamefont {Liu}},\ and\ \bibinfo {author} {\bibfnamefont {C.}~\bibnamefont {Liu}},\ }\href@noop {} {\bibinfo {title} {{Observation of Spin Splitting in Room-Temperature Metallic Antiferromagnet CrSb}}} (\bibinfo {year} {2024}),\ \Eprint {https://arxiv.org/abs/2405.12679} {arXiv:2405.12679 [cond-mat.mtrl-sci]} \BibitemShut {NoStop}%
\bibitem [{\citenamefont {Yang}\ \emph {et~al.}(2024)\citenamefont {Yang}, \citenamefont {Li}, \citenamefont {Yang}, \citenamefont {Li}, \citenamefont {Zheng}, \citenamefont {Zhu}, \citenamefont {Cao}, \citenamefont {Zhao}, \citenamefont {Zhang}, \citenamefont {Ye}, \citenamefont {Song}, \citenamefont {Hu}, \citenamefont {Yang}, \citenamefont {Shi}, \citenamefont {Yuan}, \citenamefont {Zhang}, \citenamefont {Xu},\ and\ \citenamefont {Liu}}]{CrSb_ARPES4}%
  \BibitemOpen
  \bibfield  {author} {\bibinfo {author} {\bibfnamefont {G.}~\bibnamefont {Yang}}, \bibinfo {author} {\bibfnamefont {Z.}~\bibnamefont {Li}}, \bibinfo {author} {\bibfnamefont {S.}~\bibnamefont {Yang}}, \bibinfo {author} {\bibfnamefont {J.}~\bibnamefont {Li}}, \bibinfo {author} {\bibfnamefont {H.}~\bibnamefont {Zheng}}, \bibinfo {author} {\bibfnamefont {W.}~\bibnamefont {Zhu}}, \bibinfo {author} {\bibfnamefont {S.}~\bibnamefont {Cao}}, \bibinfo {author} {\bibfnamefont {W.}~\bibnamefont {Zhao}}, \bibinfo {author} {\bibfnamefont {J.}~\bibnamefont {Zhang}}, \bibinfo {author} {\bibfnamefont {M.}~\bibnamefont {Ye}}, \bibinfo {author} {\bibfnamefont {Y.}~\bibnamefont {Song}}, \bibinfo {author} {\bibfnamefont {L.-H.}\ \bibnamefont {Hu}}, \bibinfo {author} {\bibfnamefont {L.}~\bibnamefont {Yang}}, \bibinfo {author} {\bibfnamefont {M.}~\bibnamefont {Shi}}, \bibinfo {author} {\bibfnamefont {H.}~\bibnamefont {Yuan}}, \bibinfo {author} {\bibfnamefont {Y.}~\bibnamefont {Zhang}}, \bibinfo {author} {\bibfnamefont
  {Y.}~\bibnamefont {Xu}},\ and\ \bibinfo {author} {\bibfnamefont {Y.}~\bibnamefont {Liu}},\ }\href@noop {} {\bibinfo {title} {{Three-dimensional mapping and electronic origin of large altermagnetic splitting near Fermi level in CrSb}}} (\bibinfo {year} {2024}),\ \Eprint {https://arxiv.org/abs/2405.12575} {arXiv:2405.12575 [cond-mat.mtrl-sci]} \BibitemShut {NoStop}%
\bibitem [{\citenamefont {Zhang}\ \emph {et~al.}(2024)\citenamefont {Zhang}, \citenamefont {Cheng}, \citenamefont {Yin}, \citenamefont {Liu}, \citenamefont {Deng}, \citenamefont {Qiao}, \citenamefont {Shi}, \citenamefont {Zhang}, \citenamefont {Lin}, \citenamefont {Liu}, \citenamefont {Ye}, \citenamefont {Huang}, \citenamefont {Meng}, \citenamefont {Zhang}, \citenamefont {Okuda}, \citenamefont {Shimada}, \citenamefont {Cui}, \citenamefont {Zhao}, \citenamefont {Cao}, \citenamefont {Qiao}, \citenamefont {Liu},\ and\ \citenamefont {Chen}}]{RbV2Te2O_3}%
  \BibitemOpen
  \bibfield  {author} {\bibinfo {author} {\bibfnamefont {F.}~\bibnamefont {Zhang}}, \bibinfo {author} {\bibfnamefont {X.}~\bibnamefont {Cheng}}, \bibinfo {author} {\bibfnamefont {Z.}~\bibnamefont {Yin}}, \bibinfo {author} {\bibfnamefont {C.}~\bibnamefont {Liu}}, \bibinfo {author} {\bibfnamefont {L.}~\bibnamefont {Deng}}, \bibinfo {author} {\bibfnamefont {Y.}~\bibnamefont {Qiao}}, \bibinfo {author} {\bibfnamefont {Z.}~\bibnamefont {Shi}}, \bibinfo {author} {\bibfnamefont {S.}~\bibnamefont {Zhang}}, \bibinfo {author} {\bibfnamefont {J.}~\bibnamefont {Lin}}, \bibinfo {author} {\bibfnamefont {Z.}~\bibnamefont {Liu}}, \bibinfo {author} {\bibfnamefont {M.}~\bibnamefont {Ye}}, \bibinfo {author} {\bibfnamefont {Y.}~\bibnamefont {Huang}}, \bibinfo {author} {\bibfnamefont {X.}~\bibnamefont {Meng}}, \bibinfo {author} {\bibfnamefont {C.}~\bibnamefont {Zhang}}, \bibinfo {author} {\bibfnamefont {T.}~\bibnamefont {Okuda}}, \bibinfo {author} {\bibfnamefont {K.}~\bibnamefont {Shimada}}, \bibinfo {author} {\bibfnamefont
  {S.}~\bibnamefont {Cui}}, \bibinfo {author} {\bibfnamefont {Y.}~\bibnamefont {Zhao}}, \bibinfo {author} {\bibfnamefont {G.-H.}\ \bibnamefont {Cao}}, \bibinfo {author} {\bibfnamefont {S.}~\bibnamefont {Qiao}}, \bibinfo {author} {\bibfnamefont {J.}~\bibnamefont {Liu}},\ and\ \bibinfo {author} {\bibfnamefont {C.}~\bibnamefont {Chen}},\ }\href {https://arxiv.org/abs/2407.19555} {\bibinfo {title} {Crystal-symmetry-paired spin-valley locking in a layered room-temperature antiferromagnet}} (\bibinfo {year} {2024}),\ \Eprint {https://arxiv.org/abs/2407.19555} {arXiv:2407.19555 [cond-mat.str-el]} \BibitemShut {NoStop}%
\bibitem [{\citenamefont {Jiang}\ \emph {et~al.}(2024)\citenamefont {Jiang}, \citenamefont {Hu}, \citenamefont {Bai}, \citenamefont {Song}, \citenamefont {Mu}, \citenamefont {Qu}, \citenamefont {Li}, \citenamefont {Zhu}, \citenamefont {Pi}, \citenamefont {Wei}, \citenamefont {Sun}, \citenamefont {Huang}, \citenamefont {Zheng}, \citenamefont {Peng}, \citenamefont {He}, \citenamefont {Li}, \citenamefont {Luo}, \citenamefont {Li}, \citenamefont {Chen}, \citenamefont {Li}, \citenamefont {Weng},\ and\ \citenamefont {Qian}}]{KV2Se2O}%
  \BibitemOpen
  \bibfield  {author} {\bibinfo {author} {\bibfnamefont {B.}~\bibnamefont {Jiang}}, \bibinfo {author} {\bibfnamefont {M.}~\bibnamefont {Hu}}, \bibinfo {author} {\bibfnamefont {J.}~\bibnamefont {Bai}}, \bibinfo {author} {\bibfnamefont {Z.}~\bibnamefont {Song}}, \bibinfo {author} {\bibfnamefont {C.}~\bibnamefont {Mu}}, \bibinfo {author} {\bibfnamefont {G.}~\bibnamefont {Qu}}, \bibinfo {author} {\bibfnamefont {W.}~\bibnamefont {Li}}, \bibinfo {author} {\bibfnamefont {W.}~\bibnamefont {Zhu}}, \bibinfo {author} {\bibfnamefont {H.}~\bibnamefont {Pi}}, \bibinfo {author} {\bibfnamefont {Z.}~\bibnamefont {Wei}}, \bibinfo {author} {\bibfnamefont {Y.}~\bibnamefont {Sun}}, \bibinfo {author} {\bibfnamefont {Y.}~\bibnamefont {Huang}}, \bibinfo {author} {\bibfnamefont {X.}~\bibnamefont {Zheng}}, \bibinfo {author} {\bibfnamefont {Y.}~\bibnamefont {Peng}}, \bibinfo {author} {\bibfnamefont {L.}~\bibnamefont {He}}, \bibinfo {author} {\bibfnamefont {S.}~\bibnamefont {Li}}, \bibinfo {author} {\bibfnamefont {J.}~\bibnamefont
  {Luo}}, \bibinfo {author} {\bibfnamefont {Z.}~\bibnamefont {Li}}, \bibinfo {author} {\bibfnamefont {G.}~\bibnamefont {Chen}}, \bibinfo {author} {\bibfnamefont {H.}~\bibnamefont {Li}}, \bibinfo {author} {\bibfnamefont {H.}~\bibnamefont {Weng}},\ and\ \bibinfo {author} {\bibfnamefont {T.}~\bibnamefont {Qian}},\ }\href {https://arxiv.org/abs/2408.00320} {\bibinfo {title} {{Discovery of a metallic room-temperature d-wave altermagnet KV$_2$Se$_2$O}}} (\bibinfo {year} {2024}),\ \Eprint {https://arxiv.org/abs/2408.00320} {arXiv:2408.00320 [cond-mat.mtrl-sci]} \BibitemShut {NoStop}%
\bibitem [{\citenamefont {Burlet}\ \emph {et~al.}(1997)\citenamefont {Burlet}, \citenamefont {Ressouche}, \citenamefont {Malaman}, \citenamefont {Welter}, \citenamefont {Sanchez},\ and\ \citenamefont {Vulliet}}]{MnTe2_neutron}%
  \BibitemOpen
  \bibfield  {author} {\bibinfo {author} {\bibfnamefont {P.}~\bibnamefont {Burlet}}, \bibinfo {author} {\bibfnamefont {E.}~\bibnamefont {Ressouche}}, \bibinfo {author} {\bibfnamefont {B.}~\bibnamefont {Malaman}}, \bibinfo {author} {\bibfnamefont {R.}~\bibnamefont {Welter}}, \bibinfo {author} {\bibfnamefont {J.~P.}\ \bibnamefont {Sanchez}},\ and\ \bibinfo {author} {\bibfnamefont {P.}~\bibnamefont {Vulliet}},\ }\bibfield  {title} {\bibinfo {title} {{Noncollinear magnetic structure of ${\mathrm{MnTe}}_{2}$}},\ }\href {https://doi.org/10.1103/PhysRevB.56.14013} {\bibfield  {journal} {\bibinfo  {journal} {Phys. Rev. B}\ }\textbf {\bibinfo {volume} {56}},\ \bibinfo {pages} {14013} (\bibinfo {year} {1997})}\BibitemShut {NoStop}%
\bibitem [{\citenamefont {Zhu}\ \emph {et~al.}(2024)\citenamefont {Zhu}, \citenamefont {Chen}, \citenamefont {Liu}, \citenamefont {Liu}, \citenamefont {Liu}, \citenamefont {Zha}, \citenamefont {Qu}, \citenamefont {Hong}, \citenamefont {Li}, \citenamefont {Jiang}, \citenamefont {Ma}, \citenamefont {Hao}, \citenamefont {Zhu}, \citenamefont {Liu}, \citenamefont {Zeng}, \citenamefont {Jayaram}, \citenamefont {Lenger}, \citenamefont {Ding}, \citenamefont {Mo}, \citenamefont {Tanaka}, \citenamefont {Arita}, \citenamefont {Liu}, \citenamefont {Ye}, \citenamefont {Shen}, \citenamefont {Wrachtrup}, \citenamefont {Huang}, \citenamefont {He}, \citenamefont {Qiao},\ and\ \citenamefont {Liu}}]{MnTe2_ARPES}%
  \BibitemOpen
  \bibfield  {author} {\bibinfo {author} {\bibfnamefont {Y.-P.}\ \bibnamefont {Zhu}}, \bibinfo {author} {\bibfnamefont {X.}~\bibnamefont {Chen}}, \bibinfo {author} {\bibfnamefont {X.-R.}\ \bibnamefont {Liu}}, \bibinfo {author} {\bibfnamefont {Y.}~\bibnamefont {Liu}}, \bibinfo {author} {\bibfnamefont {P.}~\bibnamefont {Liu}}, \bibinfo {author} {\bibfnamefont {H.}~\bibnamefont {Zha}}, \bibinfo {author} {\bibfnamefont {G.}~\bibnamefont {Qu}}, \bibinfo {author} {\bibfnamefont {C.}~\bibnamefont {Hong}}, \bibinfo {author} {\bibfnamefont {J.}~\bibnamefont {Li}}, \bibinfo {author} {\bibfnamefont {Z.}~\bibnamefont {Jiang}}, \bibinfo {author} {\bibfnamefont {X.-M.}\ \bibnamefont {Ma}}, \bibinfo {author} {\bibfnamefont {Y.-J.}\ \bibnamefont {Hao}}, \bibinfo {author} {\bibfnamefont {M.-Y.}\ \bibnamefont {Zhu}}, \bibinfo {author} {\bibfnamefont {W.}~\bibnamefont {Liu}}, \bibinfo {author} {\bibfnamefont {M.}~\bibnamefont {Zeng}}, \bibinfo {author} {\bibfnamefont {S.}~\bibnamefont {Jayaram}}, \bibinfo {author}
  {\bibfnamefont {M.}~\bibnamefont {Lenger}}, \bibinfo {author} {\bibfnamefont {J.}~\bibnamefont {Ding}}, \bibinfo {author} {\bibfnamefont {S.}~\bibnamefont {Mo}}, \bibinfo {author} {\bibfnamefont {K.}~\bibnamefont {Tanaka}}, \bibinfo {author} {\bibfnamefont {M.}~\bibnamefont {Arita}}, \bibinfo {author} {\bibfnamefont {Z.}~\bibnamefont {Liu}}, \bibinfo {author} {\bibfnamefont {M.}~\bibnamefont {Ye}}, \bibinfo {author} {\bibfnamefont {D.}~\bibnamefont {Shen}}, \bibinfo {author} {\bibfnamefont {J.}~\bibnamefont {Wrachtrup}}, \bibinfo {author} {\bibfnamefont {Y.}~\bibnamefont {Huang}}, \bibinfo {author} {\bibfnamefont {R.-H.}\ \bibnamefont {He}}, \bibinfo {author} {\bibfnamefont {S.}~\bibnamefont {Qiao}},\ and\ \bibinfo {author} {\bibfnamefont {C.}~\bibnamefont {Liu}, \bibfnamefont {Qihang nd~Liu}},\ }\bibfield  {title} {\bibinfo {title} {{Observation of plaid-like spin splitting in a noncoplanar antiferromagnet}},\ }\href {https://doi.org/10.1038/s41586-024-07023-w} {\bibfield  {journal} {\bibinfo  {journal}
  {Nature}\ }\textbf {\bibinfo {volume} {626}},\ \bibinfo {pages} {523} (\bibinfo {year} {2024})}\BibitemShut {NoStop}%
\bibitem [{\citenamefont {Han}\ \emph {et~al.}(2024)\citenamefont {Han}, \citenamefont {Fu}, \citenamefont {Peng}, \citenamefont {Cheng}, \citenamefont {Dai}, \citenamefont {Liu}, \citenamefont {Li}, \citenamefont {Zhang}, \citenamefont {Zhu}, \citenamefont {Bai}, \citenamefont {Zhou}, \citenamefont {Liang}, \citenamefont {Chen}, \citenamefont {Wang}, \citenamefont {Chen}, \citenamefont {Yang}, \citenamefont {Zhang}, \citenamefont {Song}, \citenamefont {Liu},\ and\ \citenamefont {Pan}}]{application_mn5si3}%
  \BibitemOpen
  \bibfield  {author} {\bibinfo {author} {\bibfnamefont {L.}~\bibnamefont {Han}}, \bibinfo {author} {\bibfnamefont {X.}~\bibnamefont {Fu}}, \bibinfo {author} {\bibfnamefont {R.}~\bibnamefont {Peng}}, \bibinfo {author} {\bibfnamefont {X.}~\bibnamefont {Cheng}}, \bibinfo {author} {\bibfnamefont {J.}~\bibnamefont {Dai}}, \bibinfo {author} {\bibfnamefont {L.}~\bibnamefont {Liu}}, \bibinfo {author} {\bibfnamefont {Y.}~\bibnamefont {Li}}, \bibinfo {author} {\bibfnamefont {Y.}~\bibnamefont {Zhang}}, \bibinfo {author} {\bibfnamefont {W.}~\bibnamefont {Zhu}}, \bibinfo {author} {\bibfnamefont {H.}~\bibnamefont {Bai}}, \bibinfo {author} {\bibfnamefont {Y.}~\bibnamefont {Zhou}}, \bibinfo {author} {\bibfnamefont {S.}~\bibnamefont {Liang}}, \bibinfo {author} {\bibfnamefont {C.}~\bibnamefont {Chen}}, \bibinfo {author} {\bibfnamefont {Q.}~\bibnamefont {Wang}}, \bibinfo {author} {\bibfnamefont {X.}~\bibnamefont {Chen}}, \bibinfo {author} {\bibfnamefont {L.}~\bibnamefont {Yang}}, \bibinfo {author} {\bibfnamefont
  {Y.}~\bibnamefont {Zhang}}, \bibinfo {author} {\bibfnamefont {C.}~\bibnamefont {Song}}, \bibinfo {author} {\bibfnamefont {J.}~\bibnamefont {Liu}},\ and\ \bibinfo {author} {\bibfnamefont {F.}~\bibnamefont {Pan}},\ }\bibfield  {title} {\bibinfo {title} {{Electrical 180° switching of Néel vector in spin-splitting antiferromagnet}},\ }\href {https://doi.org/10.1126/sciadv.adn0479} {\bibfield  {journal} {\bibinfo  {journal} {Sci. Adv.}\ }\textbf {\bibinfo {volume} {10}},\ \bibinfo {pages} {eadn0479} (\bibinfo {year} {2024})}\BibitemShut {NoStop}%
\bibitem [{\citenamefont {Zhou}\ \emph {et~al.}(2025)\citenamefont {Zhou}, \citenamefont {Cheng}, \citenamefont {Hu}, \citenamefont {Chu}, \citenamefont {Bai}, \citenamefont {Han}, \citenamefont {Liu}, \citenamefont {Pan},\ and\ \citenamefont {Song}}]{crsb_transport}%
  \BibitemOpen
  \bibfield  {author} {\bibinfo {author} {\bibfnamefont {Z.}~\bibnamefont {Zhou}}, \bibinfo {author} {\bibfnamefont {X.}~\bibnamefont {Cheng}}, \bibinfo {author} {\bibfnamefont {M.}~\bibnamefont {Hu}}, \bibinfo {author} {\bibfnamefont {R.}~\bibnamefont {Chu}}, \bibinfo {author} {\bibfnamefont {H.}~\bibnamefont {Bai}}, \bibinfo {author} {\bibfnamefont {L.}~\bibnamefont {Han}}, \bibinfo {author} {\bibfnamefont {J.}~\bibnamefont {Liu}}, \bibinfo {author} {\bibfnamefont {F.}~\bibnamefont {Pan}},\ and\ \bibinfo {author} {\bibfnamefont {C.}~\bibnamefont {Song}},\ }\bibfield  {title} {\bibinfo {title} {Manipulation of the altermagnetic order in crsb via crystal symmetry},\ }\bibfield  {journal} {\bibinfo  {journal} {Nature}\ }\href {https://doi.org/10.1038/s41586-024-08436-3} {10.1038/s41586-024-08436-3} (\bibinfo {year} {2025})\BibitemShut {NoStop}%
\bibitem [{\citenamefont {Cheong}\ and\ \citenamefont {Huang}(2024)}]{AM_noncollinear}%
  \BibitemOpen
  \bibfield  {author} {\bibinfo {author} {\bibfnamefont {S.-W.}\ \bibnamefont {Cheong}}\ and\ \bibinfo {author} {\bibfnamefont {F.-T.}\ \bibnamefont {Huang}},\ }\bibfield  {title} {\bibinfo {title} {Altermagnetism with non-collinear spins},\ }\href {https://doi.org/10.1038/s41535-024-00626-6} {\bibfield  {journal} {\bibinfo  {journal} {npj Quantum Mater.}\ }\textbf {\bibinfo {volume} {9}},\ \bibinfo {pages} {13} (\bibinfo {year} {2024})}\BibitemShut {NoStop}%
\bibitem [{\citenamefont {Gallego}\ \emph {et~al.}(2016)\citenamefont {Gallego}, \citenamefont {Perez-Mato}, \citenamefont {Elcoro}, \citenamefont {Tasci}, \citenamefont {Hanson}, \citenamefont {Momma}, \citenamefont {Aroyo},\ and\ \citenamefont {Madariaga}}]{data_gallego2016magndata}%
  \BibitemOpen
  \bibfield  {author} {\bibinfo {author} {\bibfnamefont {S.~V.}\ \bibnamefont {Gallego}}, \bibinfo {author} {\bibfnamefont {J.~M.}\ \bibnamefont {Perez-Mato}}, \bibinfo {author} {\bibfnamefont {L.}~\bibnamefont {Elcoro}}, \bibinfo {author} {\bibfnamefont {E.~S.}\ \bibnamefont {Tasci}}, \bibinfo {author} {\bibfnamefont {R.~M.}\ \bibnamefont {Hanson}}, \bibinfo {author} {\bibfnamefont {K.}~\bibnamefont {Momma}}, \bibinfo {author} {\bibfnamefont {M.~I.}\ \bibnamefont {Aroyo}},\ and\ \bibinfo {author} {\bibfnamefont {G.}~\bibnamefont {Madariaga}},\ }\bibfield  {title} {\bibinfo {title} {{{\it MAGNDATA}: towards a database of magnetic structures. I.The commensurate case}},\ }\href {https://doi.org/10.1107/S1600576716012863} {\bibfield  {journal} {\bibinfo  {journal} {J. App. Cryst.}\ }\textbf {\bibinfo {volume} {49}},\ \bibinfo {pages} {1750} (\bibinfo {year} {2016})}\BibitemShut {NoStop}%
\bibitem [{\citenamefont {Liu}\ \emph {et~al.}(2022)\citenamefont {Liu}, \citenamefont {Li}, \citenamefont {Han}, \citenamefont {Wan},\ and\ \citenamefont {Liu}}]{sp_liu2022spin}%
  \BibitemOpen
  \bibfield  {author} {\bibinfo {author} {\bibfnamefont {P.}~\bibnamefont {Liu}}, \bibinfo {author} {\bibfnamefont {J.}~\bibnamefont {Li}}, \bibinfo {author} {\bibfnamefont {J.}~\bibnamefont {Han}}, \bibinfo {author} {\bibfnamefont {X.}~\bibnamefont {Wan}},\ and\ \bibinfo {author} {\bibfnamefont {Q.}~\bibnamefont {Liu}},\ }\bibfield  {title} {\bibinfo {title} {{Spin-Group Symmetry in Magnetic Materials with Negligible Spin-Orbit Coupling}},\ }\href {https://doi.org/10.1103/PhysRevX.12.021016} {\bibfield  {journal} {\bibinfo  {journal} {Phys. Rev. X}\ }\textbf {\bibinfo {volume} {12}},\ \bibinfo {pages} {021016} (\bibinfo {year} {2022})}\BibitemShut {NoStop}%
\bibitem [{\citenamefont {Xiao}\ \emph {et~al.}(2023)\citenamefont {Xiao}, \citenamefont {Zhao}, \citenamefont {Li}, \citenamefont {Shindou},\ and\ \citenamefont {Song}}]{spingroup_1}%
  \BibitemOpen
  \bibfield  {author} {\bibinfo {author} {\bibfnamefont {Z.}~\bibnamefont {Xiao}}, \bibinfo {author} {\bibfnamefont {J.}~\bibnamefont {Zhao}}, \bibinfo {author} {\bibfnamefont {Y.}~\bibnamefont {Li}}, \bibinfo {author} {\bibfnamefont {R.}~\bibnamefont {Shindou}},\ and\ \bibinfo {author} {\bibfnamefont {Z.-D.}\ \bibnamefont {Song}},\ }\href@noop {} {\bibinfo {title} {{Spin Space Groups: Full Classification and Applications}}} (\bibinfo {year} {2023}),\ \Eprint {https://arxiv.org/abs/2307.10364} {arXiv:2307.10364 [cond-mat.mes-hall]} \BibitemShut {NoStop}%
\bibitem [{\citenamefont {Jian}\ \emph {et~al.}(2023)\citenamefont {Jian}, \citenamefont {Zheng-Xin},\ and\ \citenamefont {Chen}}]{sp_fangchen2021symmetry}%
  \BibitemOpen
  \bibfield  {author} {\bibinfo {author} {\bibfnamefont {Y.}~\bibnamefont {Jian}}, \bibinfo {author} {\bibfnamefont {L.}~\bibnamefont {Zheng-Xin}},\ and\ \bibinfo {author} {\bibfnamefont {F.}~\bibnamefont {Chen}},\ }\href@noop {} {\bibinfo {title} {{Symmetry invariants in magnetically ordered systems having weak spin-orbit coupling}}} (\bibinfo {year} {2023}),\ \Eprint {https://arxiv.org/abs/2105.12738} {arXiv:2105.12738 [cond-mat.mes-hall]} \BibitemShut {NoStop}%
\bibitem [{\citenamefont {Brinkman}\ and\ \citenamefont {Elliott}(1966)}]{spingroup_2}%
  \BibitemOpen
  \bibfield  {author} {\bibinfo {author} {\bibfnamefont {W.}~\bibnamefont {Brinkman}}\ and\ \bibinfo {author} {\bibfnamefont {R.~J.}\ \bibnamefont {Elliott}},\ }\bibfield  {title} {\bibinfo {title} {Theory of spin-space groups},\ }\href@noop {} {\bibfield  {journal} {\bibinfo  {journal} {Proceedings of the Royal Society of London. Series A. Mathematical and Physical Sciences}\ }\textbf {\bibinfo {volume} {294}},\ \bibinfo {pages} {343} (\bibinfo {year} {1966})}\BibitemShut {NoStop}%
\bibitem [{\citenamefont {Litvin}\ and\ \citenamefont {Opechowski}(1974)}]{spingroup_3}%
  \BibitemOpen
  \bibfield  {author} {\bibinfo {author} {\bibfnamefont {D.~B.}\ \bibnamefont {Litvin}}\ and\ \bibinfo {author} {\bibfnamefont {W.}~\bibnamefont {Opechowski}},\ }\bibfield  {title} {\bibinfo {title} {Spin groups},\ }\href@noop {} {\bibfield  {journal} {\bibinfo  {journal} {Physica}\ }\textbf {\bibinfo {volume} {76}},\ \bibinfo {pages} {538} (\bibinfo {year} {1974})}\BibitemShut {NoStop}%
\bibitem [{\citenamefont {Kleiner}(1966)}]{MPG_class1}%
  \BibitemOpen
  \bibfield  {author} {\bibinfo {author} {\bibfnamefont {W.~H.}\ \bibnamefont {Kleiner}},\ }\bibfield  {title} {\bibinfo {title} {Space-time symmetry of transport coefficients},\ }\href {https://doi.org/10.1103/PhysRev.142.318} {\bibfield  {journal} {\bibinfo  {journal} {Phys. Rev.}\ }\textbf {\bibinfo {volume} {142}},\ \bibinfo {pages} {318} (\bibinfo {year} {1966})}\BibitemShut {NoStop}%
\bibitem [{\citenamefont {Seemann}\ \emph {et~al.}(2015)\citenamefont {Seemann}, \citenamefont {K\"odderitzsch}, \citenamefont {Wimmer},\ and\ \citenamefont {Ebert}}]{MPG_class2}%
  \BibitemOpen
  \bibfield  {author} {\bibinfo {author} {\bibfnamefont {M.}~\bibnamefont {Seemann}}, \bibinfo {author} {\bibfnamefont {D.}~\bibnamefont {K\"odderitzsch}}, \bibinfo {author} {\bibfnamefont {S.}~\bibnamefont {Wimmer}},\ and\ \bibinfo {author} {\bibfnamefont {H.}~\bibnamefont {Ebert}},\ }\bibfield  {title} {\bibinfo {title} {Symmetry-imposed shape of linear response tensors},\ }\href {https://doi.org/10.1103/PhysRevB.92.155138} {\bibfield  {journal} {\bibinfo  {journal} {Phys. Rev. B}\ }\textbf {\bibinfo {volume} {92}},\ \bibinfo {pages} {155138} (\bibinfo {year} {2015})}\BibitemShut {NoStop}%
\bibitem [{\citenamefont {Bradley}\ and\ \citenamefont {Cracknell}(2009)}]{MPG_class3}%
  \BibitemOpen
  \bibfield  {author} {\bibinfo {author} {\bibfnamefont {C.}~\bibnamefont {Bradley}}\ and\ \bibinfo {author} {\bibfnamefont {A.}~\bibnamefont {Cracknell}},\ }\href@noop {} {\emph {\bibinfo {title} {The mathematical theory of symmetry in solids: representation theory for point groups and space groups}}}\ (\bibinfo  {publisher} {Oxford University Press},\ \bibinfo {year} {2009})\BibitemShut {NoStop}%
\bibitem [{\citenamefont {Hayami}\ \emph {et~al.}(2020)\citenamefont {Hayami}, \citenamefont {Yanagi},\ and\ \citenamefont {Kusunose}}]{sp_hayami2020}%
  \BibitemOpen
  \bibfield  {author} {\bibinfo {author} {\bibfnamefont {S.}~\bibnamefont {Hayami}}, \bibinfo {author} {\bibfnamefont {Y.}~\bibnamefont {Yanagi}},\ and\ \bibinfo {author} {\bibfnamefont {H.}~\bibnamefont {Kusunose}},\ }\bibfield  {title} {\bibinfo {title} {Bottom-up design of spin-split and reshaped electronic band structures in antiferromagnets without spin-orbit coupling: Procedure on the basis of augmented multipoles},\ }\href {https://doi.org/10.1103/PhysRevB.102.144441} {\bibfield  {journal} {\bibinfo  {journal} {Phys. Rev. B}\ }\textbf {\bibinfo {volume} {102}},\ \bibinfo {pages} {144441} (\bibinfo {year} {2020})}\BibitemShut {NoStop}%
\bibitem [{\citenamefont {Liu}\ \emph {et~al.}(2023)\citenamefont {Liu}, \citenamefont {Li}, \citenamefont {Liu},\ and\ \citenamefont {Liu}}]{liu2023universal}%
  \BibitemOpen
  \bibfield  {author} {\bibinfo {author} {\bibfnamefont {Y.}~\bibnamefont {Liu}}, \bibinfo {author} {\bibfnamefont {J.}~\bibnamefont {Li}}, \bibinfo {author} {\bibfnamefont {P.}~\bibnamefont {Liu}},\ and\ \bibinfo {author} {\bibfnamefont {Q.}~\bibnamefont {Liu}},\ }\bibfield  {title} {\bibinfo {title} {Universal theory of spin-momentum-orbital-site locking},\ }\href@noop {} {\bibfield  {journal} {\bibinfo  {journal} {arXiv preprint arXiv:2306.16312}\ } (\bibinfo {year} {2023})}\BibitemShut {NoStop}%
\bibitem [{\citenamefont {Borovik-romanov}(1994)}]{PZM_1994}%
  \BibitemOpen
  \bibfield  {author} {\bibinfo {author} {\bibfnamefont {A.~S.}\ \bibnamefont {Borovik-romanov}},\ }\bibfield  {title} {\bibinfo {title} {Piezomagnetism, linear magnetostriction and magnetooptic effect},\ }\href {https://doi.org/10.1080/00150199408245101} {\bibfield  {journal} {\bibinfo  {journal} {Ferroelectrics}\ }\textbf {\bibinfo {volume} {162}},\ \bibinfo {pages} {153} (\bibinfo {year} {1994})}\BibitemShut {NoStop}%
\bibitem [{\citenamefont {Ablimit}\ \emph {et~al.}(2018{\natexlab{a}})\citenamefont {Ablimit}, \citenamefont {Sun}, \citenamefont {Cheng}, \citenamefont {Liu}, \citenamefont {Wu}, \citenamefont {Jiang}, \citenamefont {Ren}, \citenamefont {Li},\ and\ \citenamefont {Cao}}]{RbV2Te2O_1}%
  \BibitemOpen
  \bibfield  {author} {\bibinfo {author} {\bibfnamefont {A.}~\bibnamefont {Ablimit}}, \bibinfo {author} {\bibfnamefont {Y.-L.}\ \bibnamefont {Sun}}, \bibinfo {author} {\bibfnamefont {E.-J.}\ \bibnamefont {Cheng}}, \bibinfo {author} {\bibfnamefont {Y.-B.}\ \bibnamefont {Liu}}, \bibinfo {author} {\bibfnamefont {S.-Q.}\ \bibnamefont {Wu}}, \bibinfo {author} {\bibfnamefont {H.}~\bibnamefont {Jiang}}, \bibinfo {author} {\bibfnamefont {Z.}~\bibnamefont {Ren}}, \bibinfo {author} {\bibfnamefont {S.}~\bibnamefont {Li}},\ and\ \bibinfo {author} {\bibfnamefont {G.-H.}\ \bibnamefont {Cao}},\ }\bibfield  {title} {\bibinfo {title} {{V$_2$Te$_2$O}: A two-dimensional van der waals correlated metal},\ }\href {https://doi.org/10.1021/acs.inorgchem.8b02280} {\bibfield  {journal} {\bibinfo  {journal} {Inorganic Chemistry}\ }\textbf {\bibinfo {volume} {57}},\ \bibinfo {pages} {14617} (\bibinfo {year} {2018}{\natexlab{a}})}\BibitemShut {NoStop}%
\bibitem [{\citenamefont {Ablimit}\ \emph {et~al.}(2018{\natexlab{b}})\citenamefont {Ablimit}, \citenamefont {Sun}, \citenamefont {Jiang}, \citenamefont {Wu}, \citenamefont {Liu},\ and\ \citenamefont {Cao}}]{RbV2Te2O_2}%
  \BibitemOpen
  \bibfield  {author} {\bibinfo {author} {\bibfnamefont {A.}~\bibnamefont {Ablimit}}, \bibinfo {author} {\bibfnamefont {Y.-L.}\ \bibnamefont {Sun}}, \bibinfo {author} {\bibfnamefont {H.}~\bibnamefont {Jiang}}, \bibinfo {author} {\bibfnamefont {S.-Q.}\ \bibnamefont {Wu}}, \bibinfo {author} {\bibfnamefont {Y.-B.}\ \bibnamefont {Liu}},\ and\ \bibinfo {author} {\bibfnamefont {G.-H.}\ \bibnamefont {Cao}},\ }\bibfield  {title} {\bibinfo {title} {Weak metal-metal transition in the vanadium oxytelluride {${\mathrm{Rb}}_{1\ensuremath{-}\ensuremath{\delta}}{\mathrm{V}}_{2}{\mathrm{Te}}_{2}\mathrm{O}$}},\ }\href {https://doi.org/10.1103/PhysRevB.97.214517} {\bibfield  {journal} {\bibinfo  {journal} {Phys. Rev. B}\ }\textbf {\bibinfo {volume} {97}},\ \bibinfo {pages} {214517} (\bibinfo {year} {2018}{\natexlab{b}})}\BibitemShut {NoStop}%
\bibitem [{\citenamefont {Gallego}\ \emph {et~al.}(2019)\citenamefont {Gallego}, \citenamefont {Etxebarria}, \citenamefont {Elcoro}, \citenamefont {Tasci},\ and\ \citenamefont {Perez-Mato}}]{Bilbao_tensor}%
  \BibitemOpen
  \bibfield  {author} {\bibinfo {author} {\bibfnamefont {S.~V.}\ \bibnamefont {Gallego}}, \bibinfo {author} {\bibfnamefont {J.}~\bibnamefont {Etxebarria}}, \bibinfo {author} {\bibfnamefont {L.}~\bibnamefont {Elcoro}}, \bibinfo {author} {\bibfnamefont {E.~S.}\ \bibnamefont {Tasci}},\ and\ \bibinfo {author} {\bibfnamefont {J.~M.}\ \bibnamefont {Perez-Mato}},\ }\bibfield  {title} {\bibinfo {title} {{Automatic calculation of symmetry-adapted tensors in magnetic and non-magnetic materials: a new tool of the Bilbao Crystallographic Server}},\ }\href {https://doi.org/10.1107/S2053273319001748} {\bibfield  {journal} {\bibinfo  {journal} {Acta Crystallographica Section A}\ }\textbf {\bibinfo {volume} {75}},\ \bibinfo {pages} {438} (\bibinfo {year} {2019})}\BibitemShut {NoStop}%
\bibitem [{\citenamefont {Liu}\ \emph {et~al.}(2025)\citenamefont {Liu}, \citenamefont {Dai},\ and\ \citenamefont {Bl{\"u}gel}}]{qihang2025}%
  \BibitemOpen
  \bibfield  {author} {\bibinfo {author} {\bibfnamefont {Q.}~\bibnamefont {Liu}}, \bibinfo {author} {\bibfnamefont {X.}~\bibnamefont {Dai}},\ and\ \bibinfo {author} {\bibfnamefont {S.}~\bibnamefont {Bl{\"u}gel}},\ }\bibfield  {title} {\bibinfo {title} {Different facets of unconventional magnetism},\ }\bibfield  {journal} {\bibinfo  {journal} {Nature Physics}\ }\href {https://doi.org/10.1038/s41567-024-02750-3} {10.1038/s41567-024-02750-3} (\bibinfo {year} {2025})\BibitemShut {NoStop}%
\end{thebibliography}
\end{document}